\documentclass{article} 
\usepackage[final]{colm2026_conference}

\usepackage{microtype}
\usepackage{hyperref}
\usepackage{url}
\usepackage{booktabs}

\usepackage{graphicx}
\usepackage{booktabs}
\usepackage{multirow}

\usepackage{tabularx}    
\usepackage{amsmath}    
\usepackage{amssymb}   
\usepackage{cleveref}
\usepackage{pifont}
\usepackage{tcolorbox}
\usepackage{multicol}
\usepackage{makecell}
\usepackage{wrapfig}

\usepackage{lineno}

\definecolor{darkblue}{rgb}{0, 0, 0.5}
\hypersetup{colorlinks=true, citecolor=darkblue, linkcolor=darkblue, urlcolor=darkblue}

\title{A11yn: Aligning LLMs for Web Accessibility-Aware UI \\Generation}

\author{Janghan Yoon\textsuperscript{1}\quad
     Jaegwan Cho\textsuperscript{1}\quad
     Junhyeok Kim\textsuperscript{1}\quad
     Jiwan Chung\textsuperscript{1}\\ 
     \textbf{Jaehyun Jeon\textsuperscript{1}}\quad
     \textbf{Seungwon Lim\textsuperscript{1}}\quad
     \textbf{Youngjae Yu\textsuperscript{2}}\\
     Yonsei University\textsuperscript{1} \quad Seoul National University\textsuperscript{2}\\
     \small{\href{mailto:jeffrobot99@yonsei.ac.kr}{\texttt{jeffrobot99@yonsei.ac.kr}}}
}

\begin{document}

\ifcolmsubmission
\linenumbers
\fi

\maketitle

\begin{abstract}
Large language models can generate visually coherent web UIs from natural language requests, but they frequently violate Web Content Accessibility Guidelines (WCAG), excluding users with diverse needs and contexts. We address this by introducing A11yn, a post-training framework for web accessibility-aware web UI generation. A11yn converts off-the-shelf WCAG audits into a verifiable web accessibility reward, enabling accessibility alignment without data from dense human annotations. To support training and evaluation, we release UIReq-6.8K, a dataset of 6,800 diverse UI generation instructions, and RealUIReq-300, a benchmark of 300 realistic web UI generation tasks. Experimental results show that A11yn achieves the strongest web accessibility compliance among baselines, reducing the inaccessibility rate by 87.5\% and 58.1\% under two independent WCAG auditors relative to the base model. Unlike prompting-based post-hoc accessibility-correction baselines, A11yn achieves these gains in a single-pass generation setting while maintaining comparable semantic fidelity and visual quality. The code and dataset are available at \url{https://github.com/jeffrobot/A11yn}.
\end{abstract}

\section{Introduction}



Large Language Models (LLMs) can generate web interfaces from natural language requests, and prior UI-generation work has primarily emphasized design fidelity and visual coherence~\citep{xiao2025designbenchcomprehensivebenchmarkmllmbased,zhou2025declarui,lu2025webgenbenchevaluatingllmsgenerating}. However, \textit{Web Accessibility} is often missing from the interface-quality objectives used in LLM-based web UI generation. Web accessibility is a fundamental concept that concerns whether interfaces can be perceived and operated by users with diverse abilities; for example, whether content is interpretable by screen readers for blind users or navigable via keyboard-only interaction for users with motor impairments. The World Wide Web Consortium (W3C) formalizes web accessibility requirements through the Web Content Accessibility Guidelines (WCAG)~\citep{w3c_wcag22_2024}. WCAG is a technical standard developed through a rigorous international, multi-stakeholder process involving accessibility experts, people with disabilities, and organizations, providing a shared and actionable baseline for web accessibility. Consistent with this standards-based framing, web accessibility auditors are widely used for reproducible web UI evaluation through violation detection~\citep{alsaeedi2020comparing, pool2023accessibility, fischer2025coverage}.

Despite mature WCAG standards and auditing tools, basic web accessibility failures remain pervasive: large-scale audits report detectable WCAG violations on over 90\% of public web pages~\citep{mowar2025codea11y}. LLMs trained on such web-scale data reproduce these defects in generated UIs. Recent studies consistently show missing alternative text, weak semantic landmarks, and improperly labeled controls in LLM-generated code~\citep{suh2025humanllmcomparativestudy, aljedaani2024does, gurictua2025llm}. FeedA11y~\citep{suh2025humanllmcomparativestudy} demonstrates that prompting LLMs with web accessibility feedback can mitigate some violations, but it introduces unwanted violations during refinement, underscoring the limits of post-hoc prompting alone. At the application level, \cite{mowar2025codea11y} further demonstrates that web accessibility-unaware coding assistants propagate and amplify omissions among novice developers. These findings motivate a core research question: \textit{Can we train LLMs to natively generate web UIs with minimal accessibility violations?}


To address this gap, we introduce \textbf{A11yn} (pronounced \textit{align}), the first post-training framework for aligning code-generating LLMs toward web accessibility-aware UI generation. The prompting analysis (Appendix~\ref{app:prompt_scaling}) shows that explicit accessibility constraints improve accessibility but remain suboptimal, motivating behavioral alignment beyond prompt-only or post-hoc repair. A11yn therefore converts WCAG violation reports from Axe-core~\citep{axe-core} into verifiable rewards for GRPO~\citep{shao2024deepseekmathpushinglimitsmathematical}, enabling optimization and internalization of web accessibility objectives for each request.

\begin{figure}[!t]
  \centering
  \includegraphics[width=1.0\columnwidth]{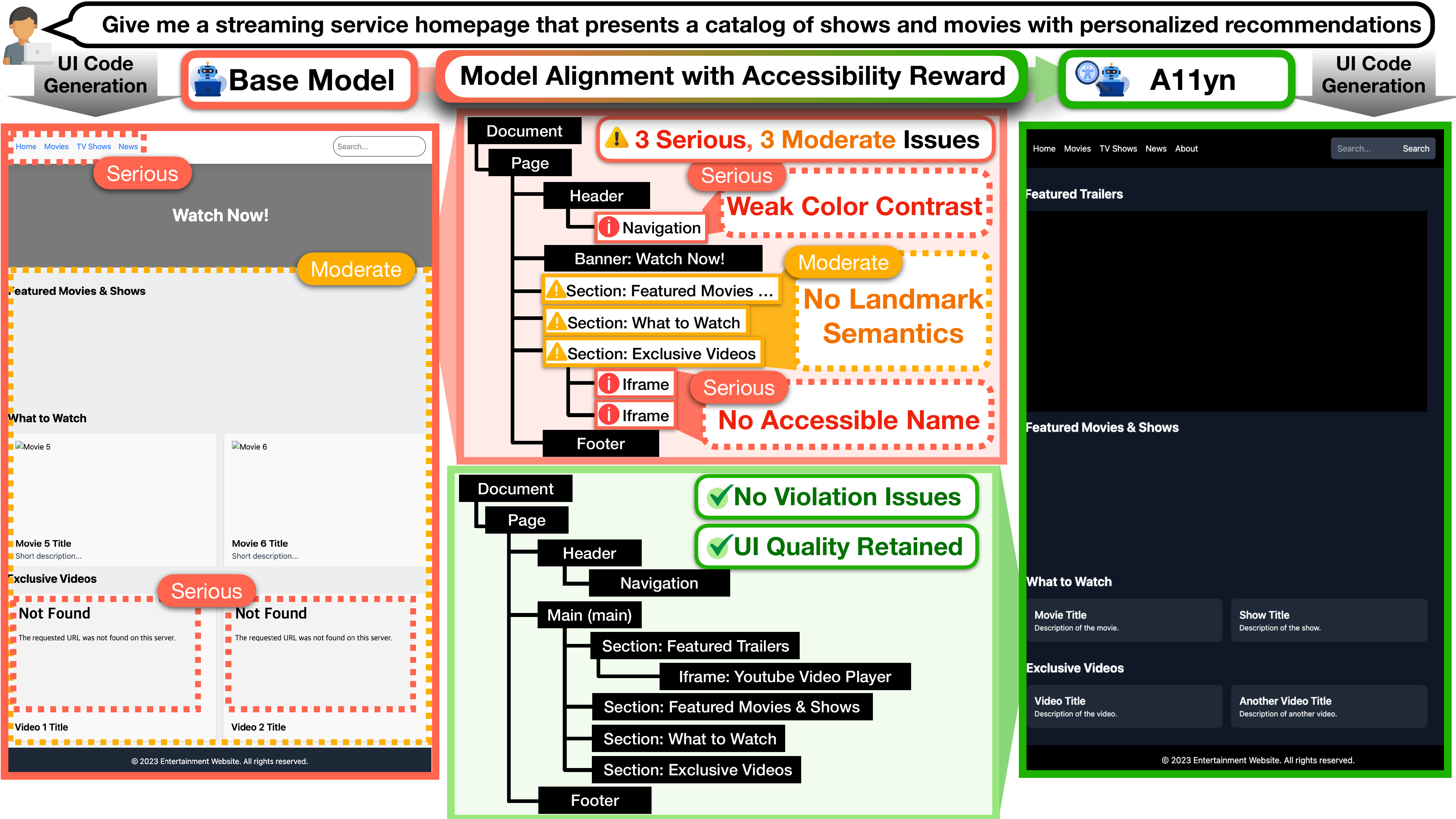}
  \caption{\textbf{Illustrative example of A11yn.} Compared with the base model, A11yn uses training-time alignment with an auditor-derived accessibility reward to generate a web UI with minimal WCAG violations in a single pass with comparable interface visual quality.}
  \label{fig:a11yn_main}
\end{figure}

Our contributions are threefold: (1) We formulate accessibility-aware UI generation as an auditor-guided reinforcement learning task and introduce a web-accessibility reward that converts off-the-shelf audit reports into a learnable signal. (2) We release two complementary resources: UIReq-6.8K for instruction-only RL training and RealUIReq-300 for realistic request-style evaluation. (3) We empirically show that A11yn improves accessibility in native generation and outperforms baselines, with gains validated across independent WCAG auditors while maintaining comparable semantic fidelity and visual quality.

\section{Related Work}
\label{sec:related}

\paragraph{LLM-based UI Code Generation.} 
With the advent of LLMs, generating UI code directly from high-level natural language descriptions has become feasible. Existing approaches have primarily focused on visual semantics, design, and functionality. UICoder~\citep{wu-2024-uicoder} improves generation quality via automated feedback during fine-tuning, WebGen-Bench~\citep{lu2025webgenbenchevaluatingllmsgenerating} benchmarks aesthetic and functional multi-page web application generation, and WebGen-Agent~\citep{lu2025webgen} enhances website generation from the perspective of visual design and functionality through post-training of LLMs. However, these methods do not explicitly optimize LLMs for web accessibility compliance.

\paragraph{Web Accessibility Assessment with Auditors}

Web accessibility evaluation with human participants is indispensable, but difficult to scale because of recruitment constraints and context-dependent variability~\citep{mack2021we}. Consequently, a broad body of HCI and accessibility research adopts WCAG-based auditors as a reproducible and objective compliance baseline, and operationalizes accessibility improvement as reducing detectable barriers, i.e., WCAG violations reported by those auditors~\citep{asok2025digital,ismail2022web, hong2007evaluating, axe-core, 10.1145/1805986.1806019, suh2025humanllmcomparativestudy, alsaeedi2020comparing, pool2023accessibility, fischer2025coverage}. Furthermore, cross-validation across multiple tools is recommended to improve assessment robustness~\citep{suh2025humanllmcomparativestudy, alsaeedi2020comparing, pool2023accessibility, fischer2025coverage}. Following the established protocol, we use automated auditors as our primary evaluation method for WCAG compliance: we train and evaluate with Axe-core and report independent cross-auditor validation with AChecker.

\paragraph{LLM-based Accessibility Enhancement for Web UIs.}
Although web accessibility is fundamental for inclusive interfaces, it remains under-optimized in LLM-based web UI generation~\citep{martins2024large, gurictua2025llm, ahmed2025codecomplianceassessingchatgpts, aljedaani2024does}. This gap has practical consequences: CodeA11y shows that accessibility-unaware coding assistants can propagate and amplify inaccessible patterns for novice developers~\citep{mowar2025codea11y}, and recent studies find that many generated UIs still fail baseline WCAG-aligned requirements~\citep{suh2025humanllmcomparativestudy, mowar2025codea11y}. Prior efforts, including FeedA11y~\citep{suh2025humanllmcomparativestudy}, a VS Code plugin~\citep{cali2025prototype}, and real-time in-DOM correction~\citep{huang2024accesspromptengineeringautomated}, improve accessibility via iterative feedback or post-hoc correction. However, new violations are introduced during refinement, motivating training-time methods that prioritize accessibility during single-pass generation~\citep{suh2025humanllmcomparativestudy}.

\begin{figure}[!t]
  \centering
  \includegraphics[width=1.0\columnwidth]{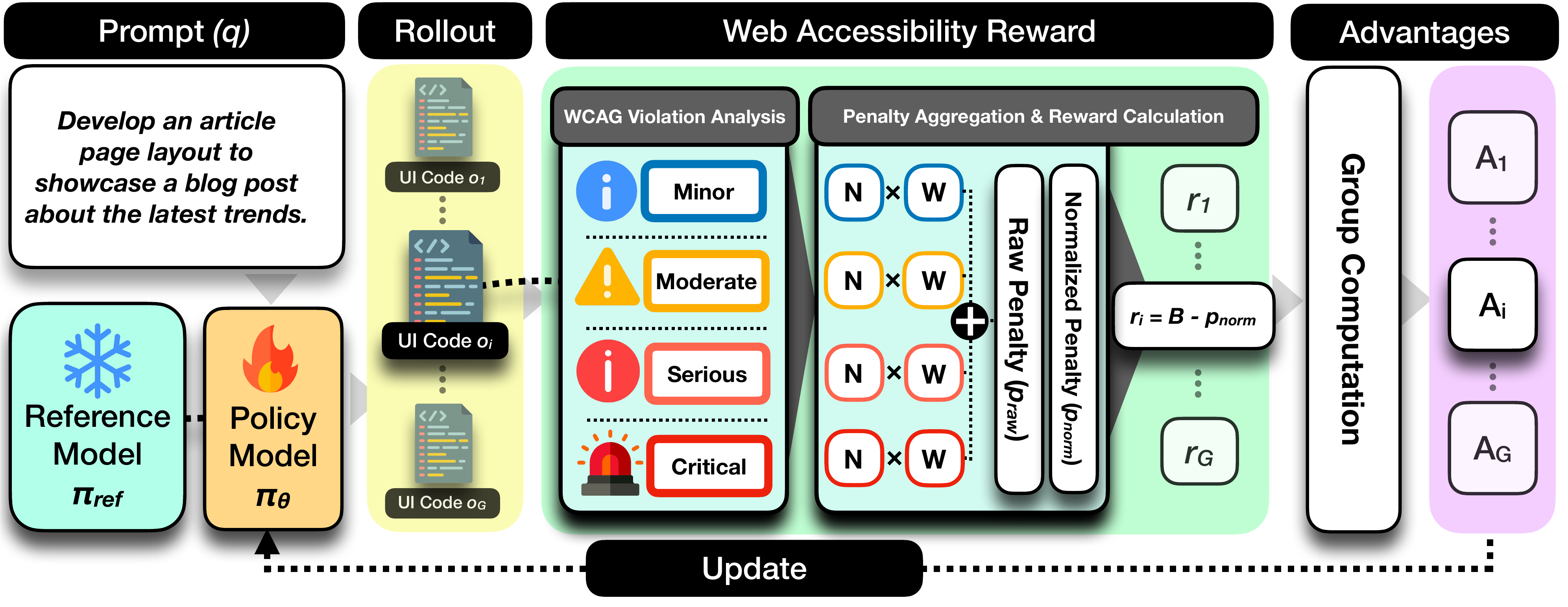}
  \caption{\textbf{A11yn aligns UI-generative LLMs using Web-Accessibility Reward.} For an instruction $q$, the policy LLM $\pi_\theta$ generates candidate UI codes $\{o_1, \ldots, o_G\}$. Off-the-shelf violation feedback from a WCAG-based accessibility auditor is converted into a penalty-aware, objectively verifiable reward signal $\{r_1, \ldots, r_G\}$ for each candidate. Rewards are normalized within the candidate set to compute advantages, which guide GRPO updates of $\pi_\theta$ toward outputs that prioritize web accessibility improvements.}
  \label{fig:a11yn_figure2}
\end{figure}

\section{Methodology}
\label{sec:method}

A11yn incorporates a novel, verifiable web accessibility reward for optimizing the LLM policy for web UI code generation. This section presents (1) the preliminary background of our approach, (2) web accessibility reward design, and (3) the overall training pipeline.

\subsection{Preliminary Background}
\label{sec:grpo}

We adopt Group Relative Policy Optimization (GRPO)~\citep{shao2024deepseekmathpushinglimitsmathematical}, a policy gradient method that optimizes policies by comparing multiple sampled completions for the same prompt. Given a UI request \(q\), the policy samples a group of candidates \(\{o_1, \dots, o_G\}\), each assigned a scalar reward \(r_i\). Rewards are normalized within the group to compute relative advantages, emphasizing comparative improvements rather than absolute scores. GRPO updates the policy using a clipped surrogate objective with KL regularization against a frozen reference model:

\begin{equation}
J_{\mathrm{GRPO}}(\theta)=\mathbb{E}_{q,\{o_i\}\sim\pi_{\theta_{\mathrm{old}}}}\!\left[\frac{1}{G}\sum_{i=1}^{G}\frac{1}{|o_i|}\sum_{t=1}^{|o_i|}L_t^{(i)}(\theta)\right]-\beta\,D_{\mathrm{KL}}(\pi_\theta\Vert\pi_{\mathrm{ref}}).
\label{eq:grpo}
\end{equation}

Unlike PPO~\citep{schulman2017proximalpolicyoptimizationalgorithms} that relies on critic-estimated absolute returns, GRPO is critic-free and uses within-prompt candidate comparisons. This makes GRPO a better match for accessibility optimization, where rewards are most informative as relative improvements among alternative UIs for the same request and are more stable under sparse, prompt-dependent signals.

\subsection{Web-Accessibility Reward}
\label{sec:web-accessibility-reward}

To train the LLM for accessible web UI generation, we design a reward that steers the model to be aware of WCAG accessibility violations. The objective is to reward outputs with minimal violations while avoiding strong incentives toward trivially simpler interfaces, which can degrade UI quality and the model's generation capability throughout training.

\paragraph{Raw Penalty.}
After the policy LLM $\pi_{\theta}$ generates each web UI output, we run Axe-core~\citep{axe-core}, a widely used open-source accessibility engine, to detect WCAG violations. For each completion, Axe-core reports affected DOM nodes together with severity levels $v \in \{\text{Minor, Moderate, Serious, Critical}\}$. For output $o_i$, $\mathcal{V}(o_i)$ denotes the set of detected severities and $N_v(o_i)$ is the number of affected nodes at severity $v$. We define the raw penalty as the severity-weighted sum of violated nodes:
\begin{equation}
P_{\mathrm{raw}}^{(i)} = \sum_{v \in \mathcal{V}(o_i)} N_v(o_i)\cdot w_v,
\end{equation}

where $w_v$ is the penalty weight assigned to severity level $v$. The ablation study in Section~\ref{sec:weightvsuniform} supports the choice of severity-dependent rather than uniform weighting.

\paragraph{DOM-Size Normalization.}
Raw penalty provides a violation-aware learning signal that encourages the policy to simply prefer samples with minimal accessibility failures. However, as larger DOMs naturally expose more opportunities for violations, optimizing raw penalty alone can bias the policy toward shorter, simpler outputs. This is especially problematic in GRPO, where candidates sampled for the same prompt are ranked within each rollout group despite potential differences in DOM size. We therefore normalize the raw penalty $P_{\mathrm{raw}}$ by DOM size as $P_{\mathrm{norm}}$, where $D_i$ is the number of rendered DOM elements in output $o_i$. This yields a size-aware accessibility penalty that enables fairer comparison among candidates with different interface sizes. We examine the empirical effect of this normalization against raw penalty in Section~\ref{ablation} and Figure~\ref{fig:a11yn-raw-vs-norm}. Finally, we define the reward by subtracting the normalized penalty from the base score $B$, which we set to 1.0.
\begin{equation}
R_i = B - P_{\mathrm{norm}}^{(i)}, \qquad \text{\textit{where}} \qquad P_{\mathrm{norm}}^{(i)} = \frac{P_{\mathrm{raw}}^{(i)}}{D_i}.
\end{equation}
Under this quantitative reward signaling scheme, a violation-free output converges toward $1.0$ as illustrated in Figure~\ref{fig:accessibility_reward} in the appendix. In practice, the policy model receives a comparable numerical score that is appropriate for giving RL feedback.

\subsection{Training Pipeline}
\label{sec:pipeline}

As illustrated in Figure~\ref{fig:a11yn_figure2}, we instantiate A11yn as a GRPO-based \textit{rollout--reward--update} pipeline. We use Qwen2.5-Coder-7B-Instruct~\citep{hui2024qwen25codertechnicalreport} as the policy model $\pi_{\theta}$ and a frozen copy as the reference model $\pi_{\mathrm{ref}}$. At each iteration, we sample a UI prompt \(q\) from UIReq-6.8K (\cref{sec:trainingset}) and generate a group of \(G\) candidate completions \(\{o_i\}_{i=1}^G\). Each completion is rendered, audited by Axe-core \citep{axe-core}, and converted to reward \(r_i\) using the reward defined in \cref{sec:web-accessibility-reward}. GRPO then normalizes rewards within the group, computes relative advantages, and updates the policy toward lower-violation outputs. Because rewards are produced online through auditor calls, the pipeline scales without manually annotated accessibility labels. At inference time, the trained policy generates accessibility-aware UI code in a single pass.

\section{Data}
\label{sec:data}

\subsection{Training: UIReq-6.8K}
\label{sec:trainingset}

\begin{wrapfigure}{l}{0.5\textwidth}
  \centering
  \includegraphics[width=0.5\textwidth]{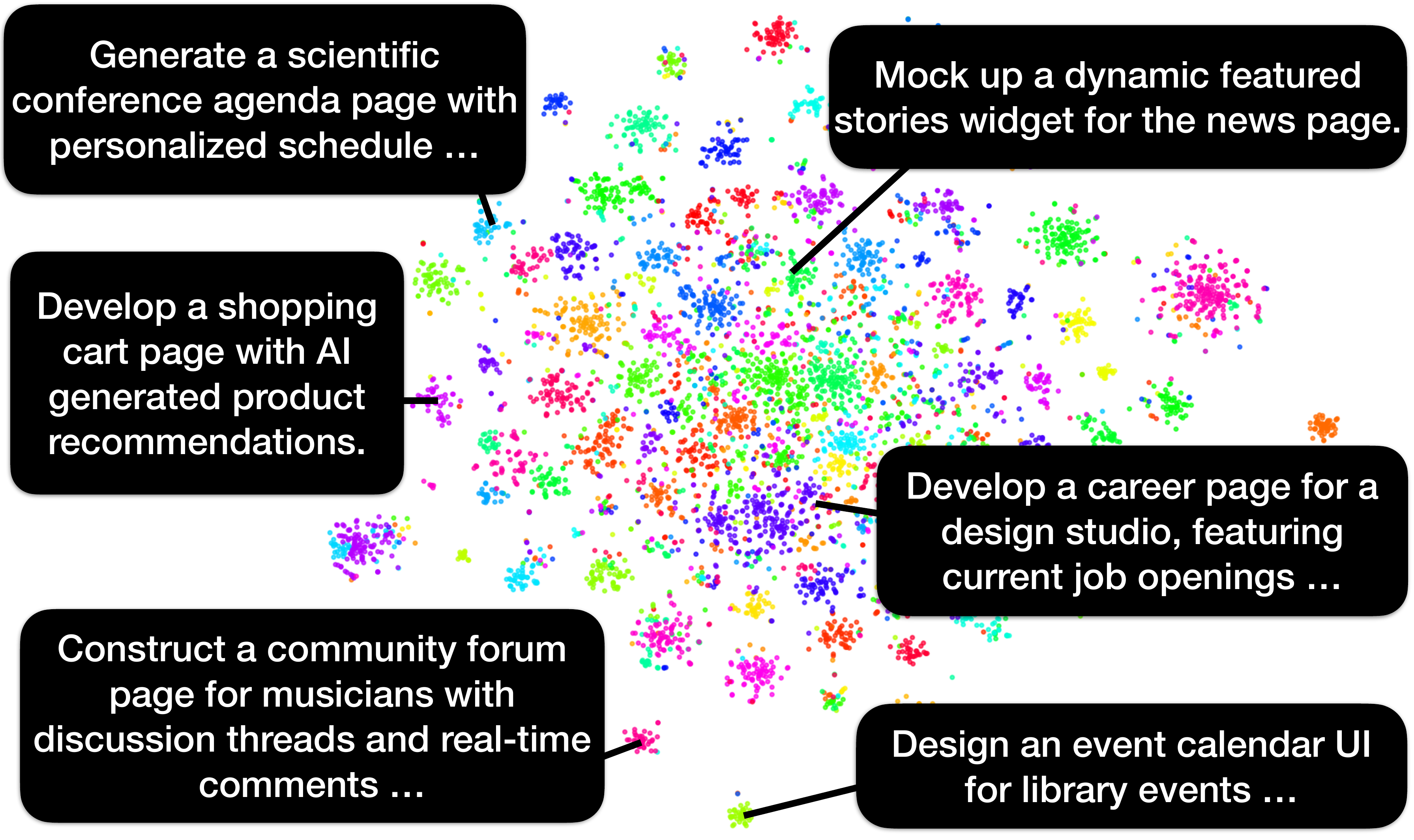}
  \caption{\textbf{t-SNE visualization of UIReq-6.8K demonstrating dataset diversity.} Each point corresponds to a UI request, colored by application category, highlighting broad coverage across domains and prompt interaction types.}
\label{fig:trainingsetdist}
\end{wrapfigure}

To train A11yn, we construct UIReq-6.8K, a reinforcement learning training dataset of 6,800 UI generation instructions. As shown in Figure~\ref{fig:trainingsetdist}, the dataset spans a wide range of domains and interaction patterns, supporting broad coverage of instructions. Unlike supervised datasets, UIReq-6.8K does not fix target UIs for each request, which enables exploration and reward optimization.

Each instruction prompt in UIReq-6.8K describes a desired user interface in natural language, specifying page type, application domain, specific web UI components, and intent (e.g. a dark-themed login screen with email and password inputs). The instruction prompts are generated using GPT-4o-mini~\citep{openai2024gpt4technicalreport} and guided to reflect diversity and semantic richness. Diversity is covered by 68 application categories (\cref{app:application_domain}). Semantic richness is enforced through detailed requirements in instruction prompt synthesis (\cref{app:prompt_detail}), where every request specifies its page type, application domain, specific UI components, and intent.

\subsection{Evaluation: RealUIReq-300}
\label{subsec:eval_set}

\begin{wrapfigure}{l}{0.5\textwidth}
  \centering
\includegraphics[width=0.5\textwidth]{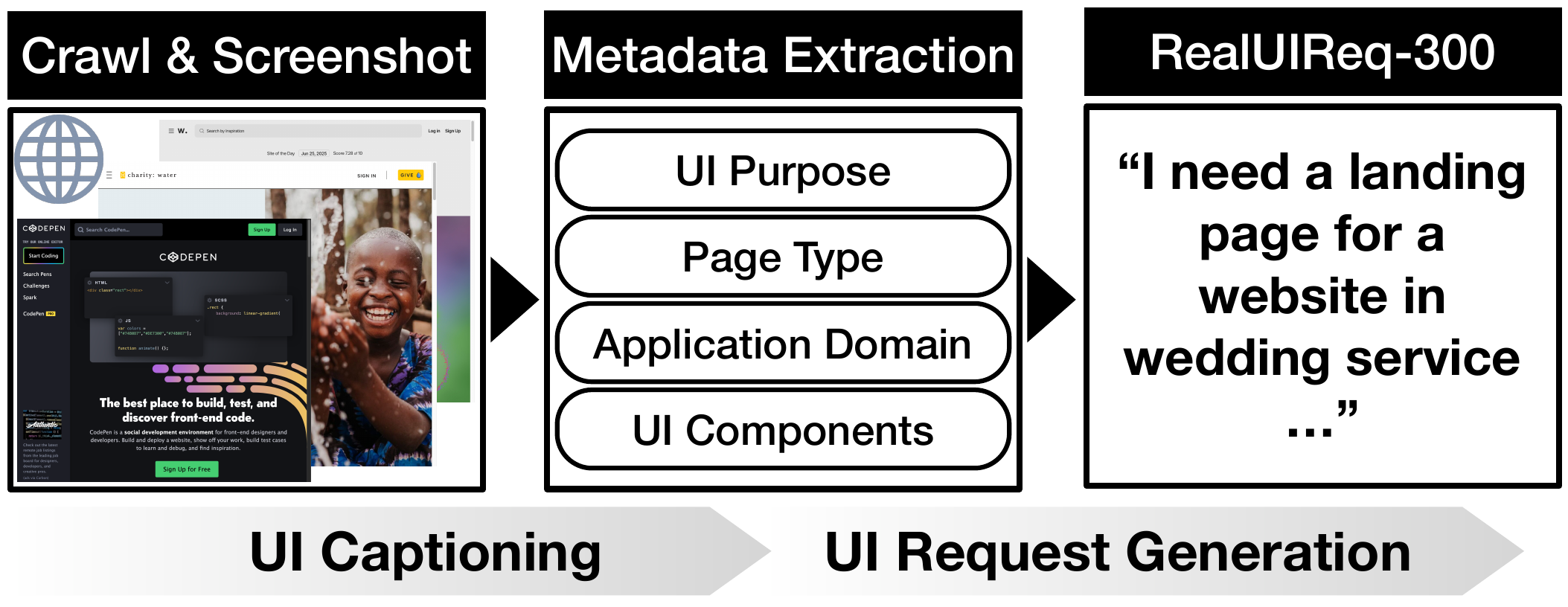}
  \caption{\textbf{RealUIReq-300 is curated from real web UIs with diverse use-cases.} UI requests are reverse-engineered from screenshots and extracted metadata, then refined into realistic instructions aligned with the original UIs.}
\label{fig:eval_data}
\vspace{-10pt}
\end{wrapfigure}

We evaluate web accessibility within a natural-language-to-web-UI code generation task, which requires request-style prompts with concrete interface details. A natural reference is Screen2Words~\citep{wang2021screen2wordsautomaticmobileui}. However, Screen2Words was introduced for UI summarization. It captures UI semantics in a narrow descriptive form: short summaries of the screen, rather than actionable requests specifying constraints.

Accessibility metrics can be applied to such open-ended prompts, but the audit scores can be confounded by variation in accessibility-relevant DOM components generated. RealUIReq-300 reduces the evaluation ambiguity through detailed, objective, real-page-grounded requests that specify expected components, enabling accessibility, semantic fidelity, and appearance to be compared under intended UI request scope. As illustrated in Figure~\ref{fig:eval_data}, each example is meticulously synthesized through a multi-stage, human-in-the-loop pipeline (\cref{data_annotation_detail}) including screenshot collection, metadata extraction, request drafting, and human filtering/refinement. GPT-4.1~\citep{openai2024gpt4technicalreport} is used as an assistive role for metadata extraction and drafting candidate requests, while human annotators review and refine the metadata and requests (Appendix~\ref{data_annotation_detail}). Accordingly, RealUIReq-300 serves as our primary benchmark, while Screen2Words remains a useful contextual reference.

\begin{table*}[h]
  \centering
  \renewcommand{\arraystretch}{1.10}  
  \setlength{\tabcolsep}{6pt}          
  \resizebox{\textwidth}{!}{
  \begin{tabular}{lcccccc}
    \toprule
    \textbf{Dataset} &
    \textbf{\makecell{Query\\Style}} &
    \textbf{\makecell{UI Intent\\Coverage}} &
    \textbf{\makecell{Component\\Details}} &
    \textbf{\makecell{UI\\Source}} &
    \textbf{\makecell{Query\\Length}} \\
    \midrule
    \makecell[l]{Screen2Words\\\citep{wang2021screen2wordsautomaticmobileui}} &
      \makecell{Single phrase,\\Summary-style} &
      \ding{55} & \ding{55} &
      \makecell{RICO dataset\\(Android UI)} &
      6 words \\
    \midrule
    \textbf{\makecell[l]{RealUIReq-300\\(Ours)}} &
      \makecell{Multi-sentence,\\Request-oriented} &
      \ding{51} & \ding{51} &
      \makecell{Real-world\\Web UI} &
      87 words \\
    \bottomrule
  \end{tabular}
  }
  \caption{\textbf{Contextual comparison of RealUIReq-300 with Screen2Words.}
  RealUIReq-300 provides multi-sentence, request-oriented instructions with structured intent and explicit component requirements grounded in real-world web UIs.}
  \label{tab:realuirequest_comparison}
\end{table*}

\section{Experimental Setup}

We evaluate A11yn and baselines by measuring the accessibility violations in their generated web UIs. In addition, we assess whether A11yn can balance accessibility with semantic alignment and visual appeal in~\cref{sec:balance}. Each model is tested on the RealUIReq-300 benchmark (\cref{subsec:eval_set}), an evaluation set of 300 web UI request prompts designed to ensure consistent and controlled comparisons. We use greedy decoding for all models to balance reproducibility with limited exploration that mitigates repetitive failure modes.

\subsection{Evaluation Metrics}

We adopt a comprehensive set of evaluation metrics informed by WCAG principles and accessibility-auditor outputs~\citep{axe-core,10.1145/1805986.1806019}. Our evaluation comprises four main metrics designed for robustness and fairness. First, we report the total number of affected-node instances across the evaluation set, categorized by Axe-core severity as \textit{Minor}, \textit{Moderate}, \textit{Serious}, and \textit{Critical}. Each instance denotes a violated DOM node reported under a specific rule, and each severity level reflects the impact of the violation, ranging from minor to critical barriers that largely hinder accessibility. Second, we use \textbf{Weighted Violation Score (WVS)}, which summarizes accessibility violations by applying severity-specific weights to affected DOM nodes. The WVS is formally defined as:

\begin{equation}
\label{wvs}
\begin{aligned}
\text{WVS}
&= \lambda_{\text{Minor}} \cdot N_{\text{Minor}}
 + \lambda_{\text{Moderate}} \cdot N_{\text{Moderate}}
 + \lambda_{\text{Serious}} \cdot N_{\text{Serious}}
 + \lambda_{\text{Critical}} \cdot N_{\text{Critical}}
\end{aligned}
\end{equation}

\noindent where $N_{\text{Minor}}$, $N_{\text{Moderate}}$, $N_{\text{Serious}}$, and $N_{\text{Critical}}$ represent the number of DOM elements with violations at each severity level for generated code on RealUIReq-300 prompts. The corresponding weights $\lambda$ reflect the relative impact of each category, with values of 1 for \textit{Minor}, 2 for \textit{Moderate}, 3 for \textit{Serious}, and 4 for \textit{Critical}. This formulation provides a single interpretable metric that captures both the frequency and severity of issues.
Third, we report the mean \textbf{Lighthouse Accessibility Score}~\citep{google2025lighthouse}, a practitioner-facing measure that aggregates automated accessibility audits from Axe-core using impact-based weights; higher values indicate better accessibility. Finally, following \cite{10.1145/355460.355549} and subsequent accessibility literature~\citep{10.1145/3377811.3380392, suh2025humanllmcomparativestudy}, we use \textbf{Inaccessibility Rate} (IR) as a conventional web accessibility measure. We adopt this normalized metric to fairly compare models that generate web content with varying lengths. Specifically, we measure IR$_{Axe}$ by calculating the ratio of WVS to the total number of DOM elements produced. In addition, as auditors may differ in WCAG rule coverage and detection heuristics, we include $\text{IR}_{\text{AChecker}}$ from \cite{suh2025humanllmcomparativestudy} for robustness, which measures inaccessibility rate based on AChecker detection heuristics.

\begin{equation}
\label{ir}
\text{IR}_{\text{Axe}} = \frac{\text{WVS}}{\text{No. of Total DOM Elements}} \qquad \text{IR}_{\text{AChecker}} = \frac{\text{No. of AChecker Violations}}{\text{No. of AChecker Flagged Elements}}
\end{equation}

\noindent These metrics capture the density of accessibility violations, enabling fair assessment and robust ranking of model candidates in web accessibility compliance.

\subsection{Baselines}

We compare A11yn with eight baselines: (1) Qwen2.5-Coder-7B-Instruct~\citep{hui2024qwen25codertechnicalreport} is the backbone from which A11yn is trained, reflecting the model's raw web UI generation capability without any accessibility optimization. (2) FeedA11y~\citep{suh2025humanllmcomparativestudy} is a direct prior effort aimed at improving web accessibility in LLM-generated UIs; here, we incorporate its prompts via a three-step iterative ReAct prompting \citep{yao2023react} loop with violation feedback from Axe-core. Implementation details are in Appendix~\ref{app:feeda11y_impl}. (3) Qwen2.5-Coder-14B-Inst is included to assess the effect of model scaling with further analysis in Appendix~\ref{app:prompt_scaling}. (4) The DPO~\citep{rafailov2023direct} baseline is trained on auditor-labeled preference pairs: for each prompt, two completions are audited, and the one with lower IR is assigned as the preferred sample. (5) The Best-of-N baseline is an inference-time method, where we sample $N$ candidates for each evaluation prompt and select the one with the lowest IR for final evaluation. We report $N=8$ for fair comparison. We also report frontier models: (6) GPT-5.1~\citep{singh2025openai}, (7) Claude Sonnet 4.6~\citep{anthropic_claude4_systemcard_2025}, and (8) Gemini 3 Pro~\citep{team2023gemini}.

\section{Results}

\begin{table*}[t]\centering\small
  \centering
  \renewcommand{\arraystretch}{1.0}
  \resizebox{\textwidth}{!}{
  \begin{tabular}{lcccccccc}
    \toprule
    \multirow{2}{*}{\textbf{Model}} &
    \multicolumn{4}{c}{\textbf{Violated DOM Counts (↓)}} &
    \multirow{2}{*}{\textbf{WVS (↓)}} &
    \multirow{2}{*}{\textbf{Lighthouse (↑)}} &
    \multicolumn{2}{c}{\textbf{IR (↓)}} \\
    \cmidrule(lr){2-5}\cmidrule(lr){8-9}
    & \textbf{Minor} & \textbf{Moderate} & \textbf{Serious} & \textbf{Critical} & & & IR$_{Axe}$ & IR$_{AChecker}$ \\
    \midrule
    \multicolumn{9}{c}{Frontier Proprietary Models} \\
    \midrule
    GPT-5.1 &
      $\mathbf{0}$ & $\underline{163}$ & $1118$ & $61$ &
      $3924$ & $92.69$ & $\underline{0.13}$ & $\underline{0.35}$\\
    Claude Sonnet 4.6 &
      $7$ & $3044$ & $4885$ & $144$ &
      $21326$ & $87.09$ & $0.45$ & $0.57$\\
    Gemini 3 Pro &
      $\underline{1}$ & $1089$ & $908$ & $57$ &
      $5131$ & $90.51$ & $0.24$ & $0.51$\\
    \midrule
    \multicolumn{9}{c}{Base Model \& Adapted Model Variants} \\
	    \midrule
	    Qwen2.5-Coder-14B-Inst &
	      $3$ & $842$ & $1652$ & $35$ &
	      $6783$ & $73.02$ & $0.48$ & $0.68$ \\
	    Qwen2.5-Coder-7B-Inst (Base) &
	      $7$ & $951$ & $794$ & $47$ &
	      $4479$ & $90.88$ & $0.40$ & $0.62$ \\
	    + FeedA11y &
	      $26$ & $1195$ & $970$ & $33$ &
	      $5458$ & $93.65$ & $0.32$ & $0.54$ \\
	    + DPO (Axe-pref.) &
	      $6$ & $1042$ & $793$ & $41$ &
	      $4633$ & $90.18$ & $0.40$ & $0.64$ \\
      + Best-of-N Sampling (N=8) &
	      $\mathbf{0}$ & $352$ & $\underline{266}$ & $\mathbf{12}$ &
	      $\underline{1550}$ & $\underline{95.43}$ & $0.14$ & $0.40$ \\
	    \midrule
	    \textbf{A11yn (Ours)} &
	      $\mathbf{0}$ & $\mathbf{74}$ &
	      $\mathbf{138}$ & $\underline{25}$ &
	      $\mathbf{662}$ & $\mathbf{97.68}$ & $\mathbf{0.05}$ &$\mathbf{0.26}$ \\
    \bottomrule
  \end{tabular}
  }
\caption{\textbf{Web accessibility measures across models.} We report Violated DOM Counts at different severity levels, Weighted Violation Score (WVS), mean Lighthouse Accessibility Score, and Inaccessibility Rate (IR). Higher is better for Lighthouse; lower is better for other metrics. \textbf{Bold} and \underline{underlined} values indicate the best and second-best values in each column, respectively.}
\label{tab:quant_results}
\end{table*}

\subsection{Quantitative Results}
Table~\ref{tab:quant_results} summarizes overall web accessibility performance of model candidates. \textbf{A11yn} achieves the strongest performance, with the lowest WVS (662), IR$_{\text{Axe}}$ (0.05), and IR$_{\text{AChecker}}$ (0.26), as well as the highest mean Lighthouse score (97.68). Relative to the base model, A11yn reduces moderate/serious/critical violations by 92.2\%, 82.6\%, and 46.8\%, respectively. Aggregate metrics show similarly strong gains: WVS drops 85.2\%, IR$_{\text{Axe}}$ drops 87.5\%, and IR$_{\text{AChecker}}$ drops 58.1\%. The underperformance of the 14B variant under default prompting (IR$_{\text{Axe}}$: 0.48 vs. 0.40; Lighthouse: 73.02 vs. 90.88) shows that scale alone does not guarantee better accessibility. However, larger models benefit more consistently when accessibility requirements are explicitly elicited through prompting (Appendix~\ref{app:prompt_scaling}). As our primary direct baseline, FeedA11y improves IR and critical counts relative to the base model, but gains are uneven with additional moderate and minor violations. FeedA11y also incurs notable inference overhead due to iterative prompting, averaging 4584 intermediate tokens per request. The DPO baseline remains close to the base model, suggesting that preference-pair optimization with the same auditor signal is less effective. Although Best-of-N yields fewer Critical affected nodes, A11yn remains stronger on aggregate metrics, showing broad accessibility gains. This residual gap is consistent with the sparsity of Critical violations in the initial training distribution (Appendix~\ref{app:training_severity}). We further extend Best-of-N up to N=20 (Appendix~\ref{app:bon_extension}), and find that larger N yields diminishing returns without closing the gap to A11yn under either auditor, suggesting that the enhancement is more than inference-time search and consistent with RL-induced policy improvement. Among frontier references, GPT-5.1 performs best, followed by Gemini 3 Pro, and Claude Sonnet 4.6.

\subsection{Analysis}

\subsubsection{WCAG Rule-Level Diagnostics}
\label{sec:wcag_rule_case_diag}

\begin{wrapfigure}{l}{0.5\columnwidth}
  \vspace{-15pt}
  \centering
  \includegraphics[width=\linewidth]{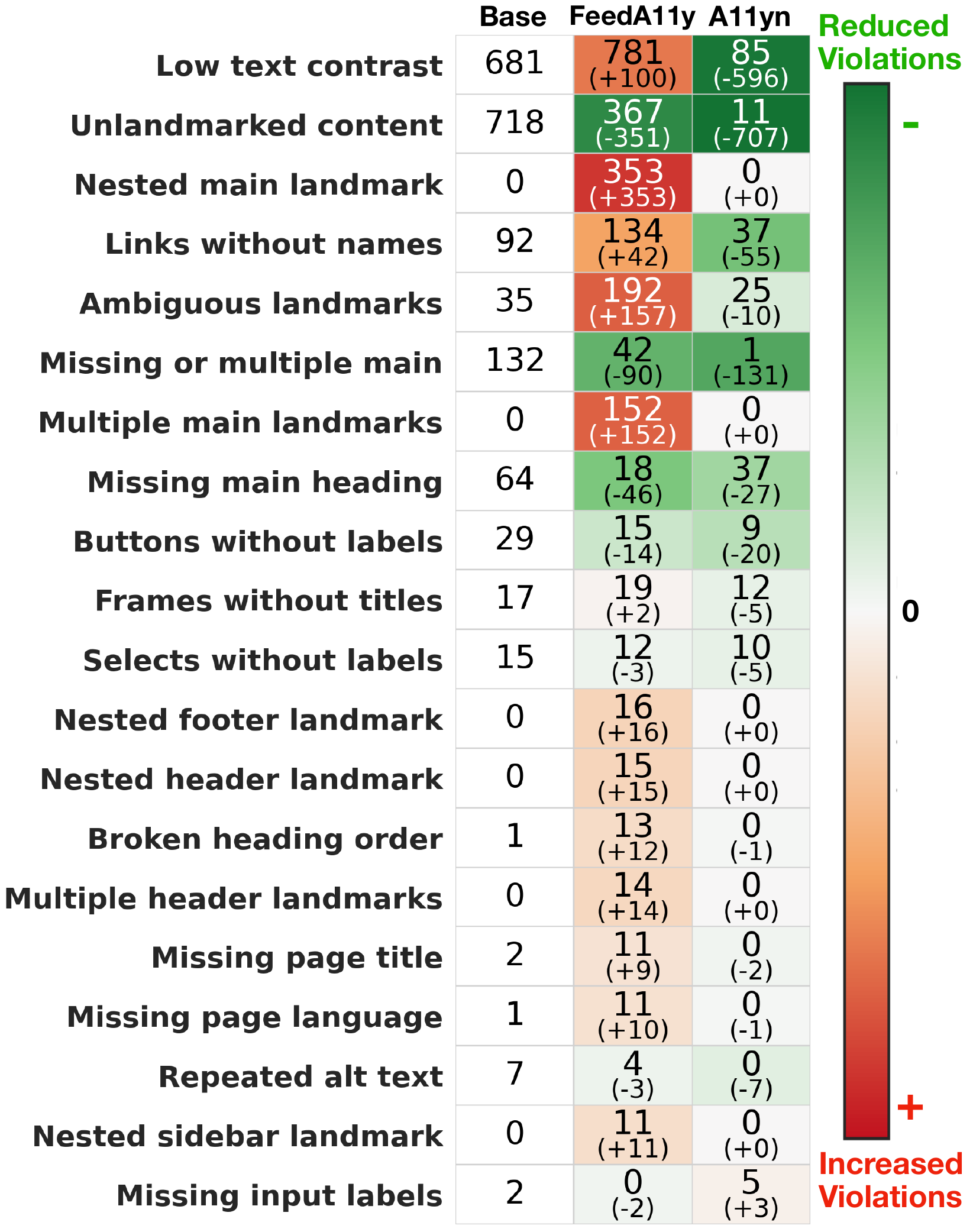}
    \caption{\textbf{WCAG rule-level diagnostics across models.} Cells show per-rule violations of Base/FeedA11y/A11yn; parentheses show change vs. Base; colors denote violation increase or reduction (Appendix ~\ref{app:wcag_rule_mapping}).}
  \label{fig:a11yn-wcag-effect}
  \vspace{-20pt}
\end{wrapfigure}

While Table~\ref{tab:quant_results} shows that FeedA11y and A11yn improve web accessibility relative to the base model, Figure~\ref{fig:a11yn-wcag-effect} provides a diagnostic view of WCAG rule analysis. Specifically, we compare FeedA11y and A11yn against the base model across key WCAG violation rules. Each cell shows the absolute violation count for a rule, with the parenthesized value and color indicating its change from the base model.

Although both methods reduce existing violation types of the base model, FeedA11y shows the inadvertent introduction of additional violations such as color-contrast and landmark-structure violations. Consistent with the finding of \cite{suh2025humanllmcomparativestudy}, post-hoc prompting remains prone to unintended regressions.

In contrast, A11yn directly addresses the failure mode flagged by FeedA11y by delivering consistent reductions across a broad range of rule categories, including major reductions in color contrast, region, and landmark-one-main. These results indicate that internalizing accessibility through training mitigates the instability of post-hoc prompting while retaining the underlying UI generation capability of the base model.
\\

\subsubsection{Preserving Semantics and Aesthetics}
\label{sec:balance}

Web UI generation is inherently multi-dimensional, requiring harmony among accessibility, semantic fidelity, and visual quality. Although accessibility and aesthetics are often framed as conflicting~\citep{anthony2019AestheticAccessibilityParadox}, prior work suggests they should be jointly optimized~\citep{kurosu1995apparent,mbipom2011interplay,le2021usability}. Accordingly, we evaluate whether A11yn's accessibility gains preserve semantic fidelity and visual appeal.

We adopt Appearance Score from WebGen-Bench~\citep{lu2025webgenbenchevaluatingllmsgenerating} with its rubric-based evaluation prompt (Appendix ~\ref{box:webgenbench}). GPT-5.1 assigns a 5-point Likert rating by jointly assessing successful rendering, content relevance, layout harmony, and modernness/beauty.

\begin{wrapfigure}{l}{0.6\columnwidth}
  \centering
  \includegraphics[width=0.6\columnwidth]{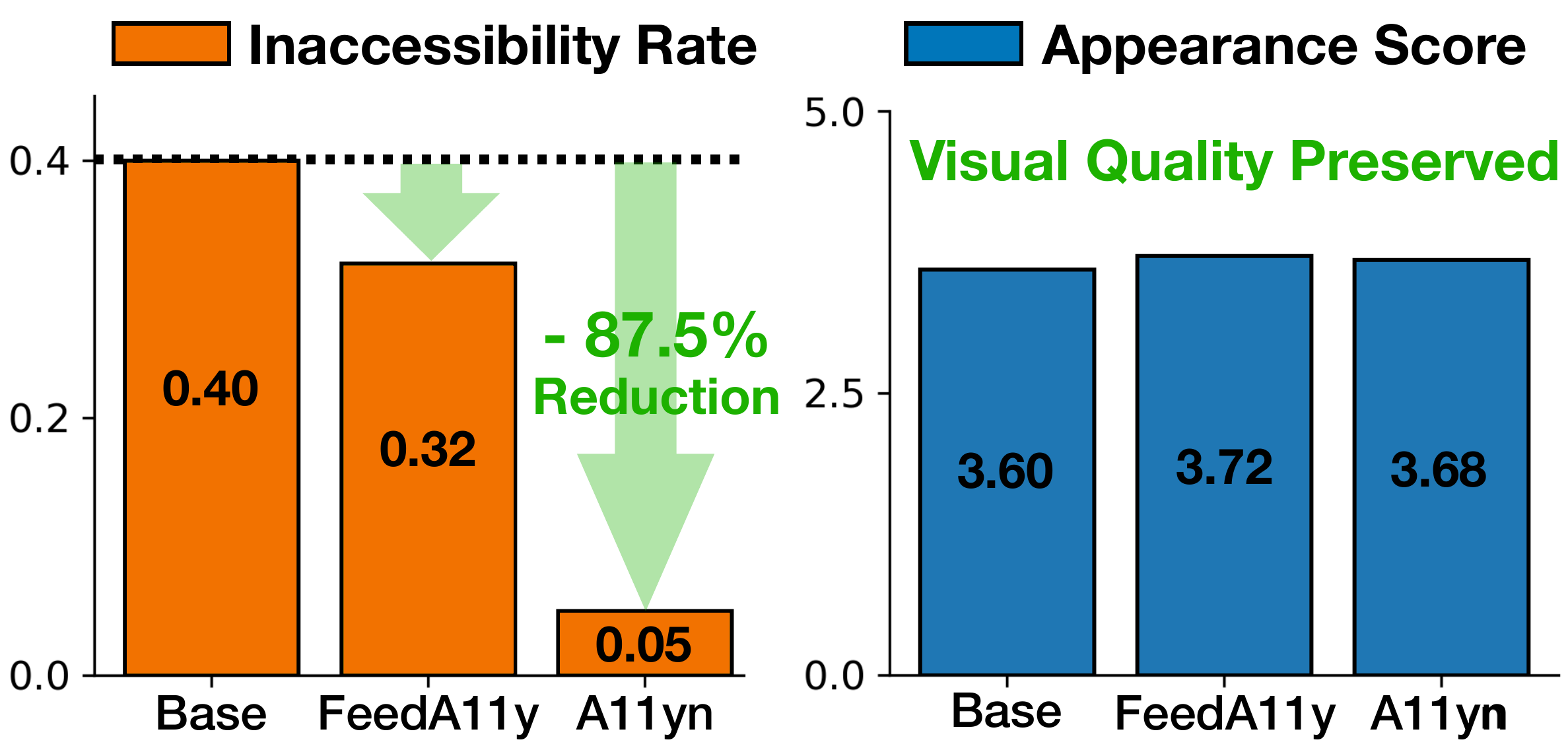}
  \caption{\textbf{Comparison of models in accessibility and visual quality.} We report Inaccessibility Rate ($\downarrow$) and Appearance Score ($\uparrow$, 5-point Likert scale).}
  \label{fig:a11yn-retain-appearancescore}
  \vspace{-15pt}
\end{wrapfigure}

Given the high correlation with human judgments reported in \cite{lu2025webgenbenchevaluatingllmsgenerating}, we treat it as a practical proxy for joint semantic-fidelity and aesthetics alignment. As shown in Figure~\ref{fig:a11yn-retain-appearancescore}, A11yn maintains appearance quality comparable to FeedA11y (3.68 vs. 3.72) while reducing inaccessibility much more strongly (0.05 vs. 0.32; 87.5\% lower than the base model at 0.40). Notably, A11yn achieves this in a single forward pass, whereas FeedA11y depends on iterative prompting.

We additionally conduct a pairwise human evaluation with five evaluators. A11yn is preferred in 39.3\% of judgments, 39.3\% rate the outputs as similar, and 21.4\% prefer the base model, supporting that A11yn preserves visual and semantic quality (Appendix~\ref{app:human_evaluation}).

\subsubsection{Ablation Study}
\label{ablation}

\begin{wraptable}{r}{0.6\columnwidth}
\vspace{-2mm}
\centering
\setlength{\tabcolsep}{2pt}
\resizebox{\linewidth}{!}{%
\begin{tabular}{lcccccc}
\toprule
\multirow{2}{*}{\textbf{Scheme}} &
\multicolumn{4}{c}{\textbf{Violated DOM Counts (↓)}} &
\multicolumn{2}{c}{\textbf{Inaccessibility Rate (↓)}} \\
\cmidrule(lr){2-5}\cmidrule(lr){6-7}
& \textbf{Minor} & \textbf{Moderate} & \textbf{Serious} & \textbf{Critical} & \textbf{IR$_{\text{Axe}}$} & \textbf{IR$_{\text{AChecker}}$} \\
\midrule
Base      & 7 & 951 & 794 & 47 & 0.40 & 0.62 \\
Uniform   & \underline{3} & \underline{126} & \underline{201} & \textbf{22} & \underline{0.07} & \underline{0.31} \\
\midrule
Weighted   & \textbf{0} & \textbf{74} & \textbf{138} & \underline{25} & \textbf{0.05} & \textbf{0.26} \\
\bottomrule
\end{tabular}
}
\caption{\label{tab:reward-ablation}
\textbf{Ablation of reward weighting strategies}. \textit{Weighted} penalties more effectively reduce WCAG violations than \textit{uniform} weighting.}
\end{wraptable}

\paragraph{Weighted vs Uniform Penalty}
\label{sec:weightvsuniform}
To evaluate the impact of severity weighting, we compare two trained A11yn variants: a \textit{weighted} reward that penalizes WCAG violations by severity and a \textit{uniform} reward that assigns equal penalties. The weighted variant uses coefficients \((0.1, 0.2, 0.3, 0.4)\) for Minor, Moderate, Serious, and Critical violations, while the uniform variant uses \((0.25, 0.25, 0.25, 0.25)\). As shown in Table~\ref{tab:reward-ablation}, severity-weighted rewards reduce accessibility violations more effectively than uniform weighting.


\begin{wrapfigure}{r}{0.5\columnwidth}
  \centering
  \includegraphics[width=0.5\columnwidth]{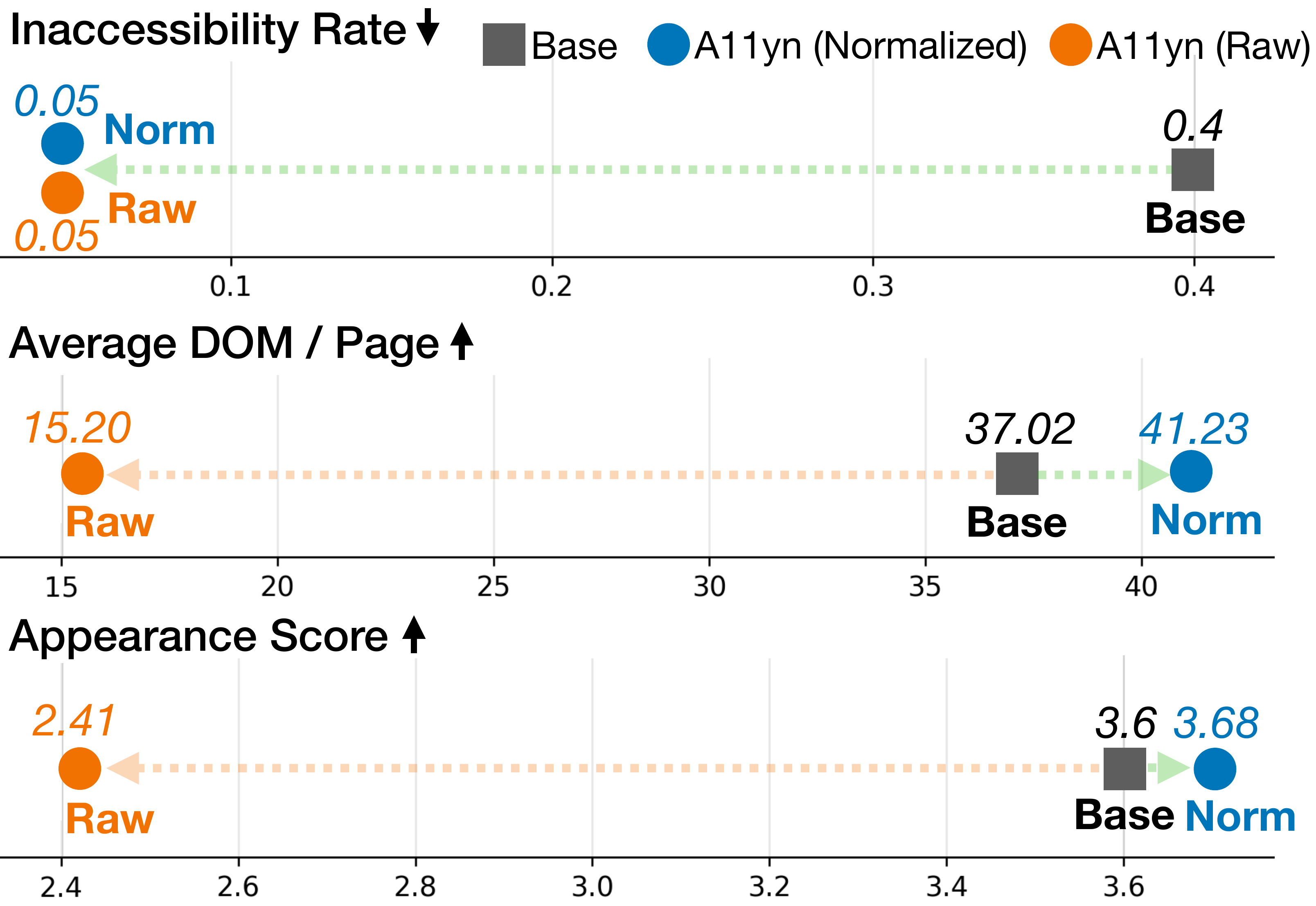}
  \caption{\textbf{Comparison of A11yn variants under Raw vs. Normalized Penalties.}
  Although A11yn trained under $P_{\text{norm}}$ and $P_{\text{raw}}$ reach similarly low inaccessibility, A11yn with a normalized penalty ($P_{\text{norm}}$) maintains UI richness and appearance.}
  \label{fig:a11yn-raw-vs-norm}
\end{wrapfigure}

\paragraph{Effect of DOM-Size Normalization} Figure~\ref{fig:a11yn-raw-vs-norm} compares two A11yn variants trained with different penalty formulations: a raw penalty ($P_{\text{raw}}$), and a DOM-normalized penalty, ($P_{\text{norm}}$). Both variants substantially improve accessibility over the base model, reducing $IR_{\text{Axe}}$ from 0.40 to 0.05. However, A11yn trained with $P_{\text{raw}}$ exhibits clear reward-hacking behavior, collapsing to much shorter outputs (average DOM/page: 15.20 vs. 37.02 for the base model), which coincides with a sharp drop in appearance score (3.60 to 2.41). In contrast, $P_{\text{norm}}$ preserves interface complexity while maintaining the same accessibility gain, producing substantially larger pages and a higher appearance score (3.68). These results show that DOM-size normalization prevents output size collapse and enables accessibility improvements without sacrificing UI richness or visual quality.



\section{Conclusion}

Web accessibility is essential yet still under-optimized in LLM-based web UI generation. We present \textbf{A11yn}, an auditor-guided RL framework that aligns code-generating LLMs with a penalty-aware, DOM-normalized WCAG reward derived from off-the-shelf accessibility audits. Enabling accessibility signals into the training objective, A11yn prioritizes web-accessible behavior at generation time.

Across our experiments, A11yn improves web accessibility compliance while preserving semantic fidelity and visual quality, with enhancements validated by Axe-core, AChecker, and Lighthouse. Relative to prompting approaches such as FeedA11y, A11yn internalizes web accessibility requirements into the policy and strengthens single-pass generation. More broadly, our findings provide evidence that verifier-checkable standards can be operationalized as practical alignment objectives for software-generation models, particularly in accessibility-aware web UI generation.

\section*{Ethics Statement}
This work aims to improve digital equity by aligning code-generating LLMs to produce accessibility-aware web UIs, thereby complying with WCAG and helping reduce barriers for users with disabilities. All training data were synthetically generated through controlled prompting, and evaluation data were curated from publicly available web pages with manual refinement; sensitive or personally identifiable content was removed. Our human evaluation involved five recruited evaluators who reviewed only generated UI screenshots and request metadata; no sensitive, personally identifiable, or private user data were collected. While misuse could enable mass generation of low-quality web pages, we mitigate this risk by committing to the open release of data, code, and documentation to guide responsible, accessibility-focused research.

\section*{Reproducibility Statement}
We ensure reproducibility by documenting all datasets, training details, and evaluation procedures. Training data (UIReq-6.8K) and the evaluation benchmark (RealUIReq-300) are fully described, along with synthesis prompts. Model training used Qwen2.5-Coder-7B-Instruct with Group-Relative Policy Optimization, with detailed hyperparameters and hardware setup provided in Section~\ref{app:train_config}. Accessibility compliance was measured using the open-source Axe-core engine distributed under the Mozilla Public License 2.0 and AChecker \citep{axe-core, 10.1145/1805986.1806019}, distributed under the GNU General Public License version 2.0 (GPLv2) to detect WCAG violations and compute reward signals. We release the code and data at \url{https://github.com/jeffrobot/A11yn}.

\section*{Use of LLMs}
For transparency, we disclose the following uses of LLMs in this work: (1) \textbf{Data generation and curation support}: LLM prompting was used to synthesize UIReq-6.8K requests and draft metadata/query candidates for RealUIReq-300, followed by author review and manual refinement; (2) \textbf{LLM-based evaluation}: consistent with the protocol of \cite{lu2025webgenbenchevaluatingllmsgenerating}, MLLM was used to assign WebGen-Bench Appearance Scores.

\section*{Acknowledgments}
This work was partly supported by the Institute of Information \& Communications Technology Planning \& Evaluation (IITP) grant funded by the Korean government (MSIT) (No. RS-2021-II211343, Artificial Intelligence Graduate School Program (Seoul National University), No. RS-2026-25522885, Development of a World Foundation Model for Training and Development of Physical AI Systems, No. RS-2026-25512061, Development of Large Action Model-Based Autonomous Digital Twin Operation Technology for Proactive Problem Solving), the National Research Foundation of Korea (NRF) grant funded by the Korean government (MSIT) (No. RS-2024-00354218, No. RS-2026-25561904), and the Technology Innovation Program (RS-2025-25456760, Development of a humanoid robot specialized in chemical processes based on AI foundation model) funded by the Ministry of Trade, Industry and Resources (MOTIR, Korea). We express special thanks to KAIT GPU project. The ICT at Seoul National University provides research facilities for this study.

\bibliography{colm2026_conference}

@misc{openai2024gpt4technicalreport,
      title={GPT-4 Technical Report}, 
      author={OpenAI and Josh Achiam and Steven Adler and Sandhini Agarwal and Lama Ahmad and Ilge Akkaya and Florencia Leoni Aleman and Diogo Almeida and Janko Altenschmidt and Sam Altman and Shyamal Anadkat and others},
      year={2024},
      eprint={2303.08774},
      archivePrefix={arXiv},
      primaryClass={cs.CL},
      url={https://arxiv.org/abs/2303.08774}, 
}

@article{team2023gemini,
  title={Gemini: a family of highly capable multimodal models},
  author={Team, Gemini and Anil, Rohan and Borgeaud, Sebastian and Alayrac, Jean-Baptiste and Yu, Jiahui and Soricut, Radu and Schalkwyk, Johan and Dai, Andrew M and Hauth, Anja and Millican, Katie and others},
  journal={arXiv preprint arXiv:2312.11805},
  year={2023}
}

@inproceedings{wu-2024-uicoder,
    title = "{UIC}oder: Finetuning Large Language Models to Generate User Interface Code through Automated Feedback",
    author = "Wu, Jason  and
      Schoop, Eldon  and
      Leung, Alan  and
      Barik, Titus  and
      Bigham, Jeffrey  and
      Nichols, Jeffrey",
    editor = "Duh, Kevin  and
      Gomez, Helena  and
      Bethard, Steven",
    booktitle = "Proceedings of the 2024 Conference of the North American Chapter of the Association for Computational Linguistics: Human Language Technologies (Volume 1: Long Papers)",
    month = jun,
    year = "2024",
    address = "Mexico City, Mexico",
    publisher = "Association for Computational Linguistics",
    url = "https://aclanthology.org/2024.naacl-long.417/",
    doi = "10.18653/v1/2024.naacl-long.417",
    pages = "7511--7525"
}

@misc{schulman2017proximalpolicyoptimizationalgorithms,
      title={Proximal Policy Optimization Algorithms}, 
      author={John Schulman and Filip Wolski and Prafulla Dhariwal and Alec Radford and Oleg Klimov},
      year={2017},
      eprint={1707.06347},
      archivePrefix={arXiv},
      primaryClass={cs.LG},
      url={https://arxiv.org/abs/1707.06347}, 
}

@article{rafailov2023direct,
  title={Direct preference optimization: Your language model is secretly a reward model},
  author={Rafailov, Rafael and Sharma, Archit and Mitchell, Eric and Manning, Christopher D and Ermon, Stefano and Finn, Chelsea},
  journal={Advances in neural information processing systems},
  volume={36},
  pages={53728--53741},
  year={2023}
}

@misc{shao2024deepseekmathpushinglimitsmathematical,
      title={DeepSeekMath: Pushing the Limits of Mathematical Reasoning in Open Language Models}, 
      author={Zhihong Shao and Peiyi Wang and Qihao Zhu and Runxin Xu and Junxiao Song and Xiao Bi and Haowei Zhang and Mingchuan Zhang and Y. K. Li and Y. Wu and Daya Guo},
      year={2024},
      eprint={2402.03300},
      archivePrefix={arXiv},
      primaryClass={cs.CL},
      url={https://arxiv.org/abs/2402.03300}, 
}

@misc{lu2025webgenbenchevaluatingllmsgenerating,
      title={WebGen-Bench: Evaluating LLMs on Generating Interactive and Functional Websites from Scratch}, 
      author={Zimu Lu and Yunqiao Yang and Houxing Ren and Haotian Hou and Han Xiao and Ke Wang and Weikang Shi and Aojun Zhou and Mingjie Zhan and Hongsheng Li},
      year={2025},
      eprint={2505.03733},
      archivePrefix={arXiv},
      primaryClass={cs.CL},
      url={https://arxiv.org/abs/2505.03733}, 
}

@inproceedings{gurictua2025llm,
  title={When LLM-Generated Code Perpetuates User Interface Accessibility Barriers, How Can We Break the Cycle},
  author={Guri{\c{t}}{\u{a}}, Alexandra-Elena and Vatavu, Radu-Daniel},
  booktitle={Proceedings of the 22nd International Web for All Conference (W4A’25)},
  year={2025}
}

@inproceedings{mowar2025codea11y,
  title={CodeA11y: Making AI Coding Assistants Useful for Accessible Web Development},
  author={Mowar, Peya and Peng, Yi-Hao and Wu, Jason and Steinfeld, Aaron and Bigham, Jeffrey P},
  booktitle={Proceedings of the 2025 CHI Conference on Human Factors in Computing Systems},
  pages={1--15},
  year={2025}
}

@misc{suh2025humanllmcomparativestudy,
      title={Human or LLM? A Comparative Study on Accessible Code Generation Capability}, 
      author={Hyunjae Suh and Mahan Tafreshipour and Sam Malek and Iftekhar Ahmed},
      year={2025},
      eprint={2503.15885},
      archivePrefix={arXiv},
      primaryClass={cs.SE},
      url={https://arxiv.org/abs/2503.15885}, 
}

@inproceedings{yao2023react,
  title={React: Synergizing reasoning and acting in language models},
  author={Yao, Shunyu and Zhao, Jeffrey and Yu, Dian and Du, Nan and Shafran, Izhak and Narasimhan, Karthik and Cao, Yuan},
  booktitle={International Conference on Learning Representations (ICLR)},
  year={2023}
}

@article{zhou2025declarui,
  title={DeclarUI: Bridging Design and Development with Automated Declarative UI Code Generation},
  author={Zhou, Ting and Zhao, Yanjie and Hou, Xinyi and Sun, Xiaoyu and Chen, Kai and Wang, Haoyu},
  journal={Proceedings of the ACM on Software Engineering},
  volume={2},
  number={FSE},
  pages={219--241},
  year={2025},
  publisher={ACM New York, NY, USA}
}

@misc{xiao2025designbenchcomprehensivebenchmarkmllmbased,
      title={DesignBench: A Comprehensive Benchmark for MLLM-based Front-end Code Generation}, 
      author={Jingyu Xiao and Ming Wang and Man Ho Lam and Yuxuan Wan and Junliang Liu and Yintong Huo and Michael R. Lyu},
      year={2025},
      eprint={2506.06251},
      archivePrefix={arXiv},
      primaryClass={cs.SE},
      url={https://arxiv.org/abs/2506.06251}, 
}

@misc{ahmed2025codecomplianceassessingchatgpts,
      title={From Code to Compliance: Assessing ChatGPT's Utility in Designing an Accessible Webpage -- A Case Study}, 
      author={Ammar Ahmed and Margarida Fresco and Fredrik Forsberg and Hallvard Grotli},
      year={2025},
      eprint={2501.03572},
      archivePrefix={arXiv},
      primaryClass={cs.HC},
      url={https://arxiv.org/abs/2501.03572}, 
}

@inproceedings{aljedaani2024does,
  title={Does chatgpt generate accessible code? investigating accessibility challenges in llm-generated source code},
  author={Aljedaani, Wajdi and Habib, Abdulrahman and Aljohani, Ahmed and Eler, Marcelo and Feng, Yunhe},
  booktitle={Proceedings of the 21st International Web for All Conference},
  pages={165--176},
  year={2024}
}

@misc{huang2024accesspromptengineeringautomated,
      title={ACCESS: Prompt Engineering for Automated Web Accessibility Violation Corrections}, 
      author={Calista Huang and Alyssa Ma and Suchir Vyasamudri and Eugenie Puype and Sayem Kamal and Juan Belza Garcia and Salar Cheema and Michael Lutz},
      year={2024},
      eprint={2401.16450},
      archivePrefix={arXiv},
      primaryClass={cs.HC},
      url={https://arxiv.org/abs/2401.16450}, 
}

@article{martins2024large,
  title={A large-scale web accessibility analysis considering technology adoption},
  author={Martins, Beatriz and Duarte, Carlos},
  journal={Universal Access in the Information Society},
  volume={23},
  number={4},
  pages={1857--1872},
  year={2024},
  publisher={Springer}
}

@inproceedings{cali2025prototype,
  title={A Prototype VS Code Extension to Improve Web Accessible Development},
  author={Cal{\`\i}, Elisa and Fulcini, Tommaso and Coppola, Riccardo and Laudadio, Lorenzo and Torchiano, Marco},
  booktitle={2025 IEEE/ACM Second IDE Workshop (IDE)},
  pages={52--57},
  year={2025},
  organization={IEEE}
}

@inproceedings{le2021usability,
  title={Usability and aesthetics: Better together for automated repair of web pages},
  author={Le-Cong, Thanh and Le, Xuan Bach D and Huynh, Quyet Thang and Le Nguyen, Phi},
  booktitle={2021 IEEE 32nd International Symposium on Software Reliability Engineering (ISSRE)},
  pages={173--183},
  year={2021},
  organization={IEEE}
}

@inproceedings{kurosu1995apparent,
  title={Apparent usability vs. inherent usability: experimental analysis on the determinants of the apparent usability},
  author={Kurosu, Masaaki and Kashimura, Kaori},
  booktitle={Conference companion on Human factors in computing systems},
  pages={292--293},
  year={1995}
}

@inproceedings{mbipom2011interplay,
  title={The interplay between web aesthetics and accessibility},
  author={Mbipom, Grace and Harper, Simon},
  booktitle={The proceedings of the 13th international ACM SIGACCESS conference on Computers and accessibility},
  pages={147--154},
  year={2011}
}

@misc{hui2024qwen25codertechnicalreport,
      title={Qwen2.5-Coder Technical Report}, 
      author={Binyuan Hui and Jian Yang and Zeyu Cui and Jiaxi Yang and Dayiheng Liu and Lei Zhang and Tianyu Liu and Jiajun Zhang and Bowen Yu and Keming Lu and Kai Dang and Yang Fan and Yichang Zhang and An Yang and Rui Men and Fei Huang and Bo Zheng and Yibo Miao and Shanghaoran Quan and Yunlong Feng and Xingzhang Ren and Xuancheng Ren and Jingren Zhou and Junyang Lin},
      year={2024},
      eprint={2409.12186},
      archivePrefix={arXiv},
      primaryClass={cs.CL},
      url={https://arxiv.org/abs/2409.12186}, 
}

@misc{axe-core,
  author = {{Deque Systems}},
  title = {{axe-core Accessibility Engine}},
  year = {2015},
  howpublished = {\url{https://github.com/dequelabs/axe-core}},
  note = {Accessed: 2025-07-28}
}

@techreport{anthropic_claude4_systemcard_2025,
  author = {Anthropic},
  title = {Claude Opus 4 \& Claude Sonnet 4: System Card},
  institution = {Anthropic},
  month = may,
  year = {2025},
  url = {https://www-cdn.anthropic.com/07b2a3f9902ee19fe39a36ca638e5ae987bc64dd.pdf},
  note = {Accessed: 2025-07-28}
}

@misc{anthony2019AestheticAccessibilityParadox,
  author       = {Anthony},
  title        = {The Aesthetic-Accessibility Paradox},
  howpublished = {\url{https://uxmovement.com/thinking/the-aesthetic-accessibility-paradox/}},
  year         = {2019},
  month        = {November},
  note         = {Accessed: 2025-07-29}
}

@misc{wang2021screen2wordsautomaticmobileui,
      title={Screen2Words: Automatic Mobile UI Summarization with Multimodal Learning}, 
      author={Bryan Wang and Gang Li and Xin Zhou and Zhourong Chen and Tovi Grossman and Yang Li},
      year={2021},
      eprint={2108.03353},
      archivePrefix={arXiv},
      primaryClass={cs.HC},
      url={https://arxiv.org/abs/2108.03353}, 
}

@inproceedings{10.1145/1805986.1806019,
author = {Gay, Greg and Li, Cindy Qi},
title = {AChecker: open, interactive, customizable, web accessibility checking},
year = {2010},
isbn = {9781450300452},
publisher = {Association for Computing Machinery},
address = {New York, NY, USA},
url = {https://doi.org/10.1145/1805986.1806019},
doi = {10.1145/1805986.1806019},
booktitle = {Proceedings of the 2010 International Cross Disciplinary Conference on Web Accessibility (W4A)},
articleno = {23},
numpages = {2},
location = {Raleigh, North Carolina},
series = {W4A '10}
}

@article{lu2025webgen,
  title={Webgen-agent: Enhancing interactive website generation with multi-level feedback and step-level reinforcement learning},
  author={Lu, Zimu and Ren, Houxing and Yang, Yunqiao and Wang, Ke and Zong, Zhuofan and Pan, Junting and Zhan, Mingjie and Li, Hongsheng},
  journal={arXiv preprint arXiv:2509.22644},
  year={2025}
}

@article{alsaeedi2020comparing,
  title={Comparing web accessibility evaluation tools and evaluating the accessibility of webpages: proposed frameworks},
  author={Alsaeedi, Abdullah},
  journal={Information},
  volume={11},
  number={1},
  pages={40},
  year={2020},
  publisher={MDPI}
}

@article{fischer2025coverage,
  title={Coverage of web accessibility guidelines provided by automated checking tools},
  author={Fischer, Thomas and Lundell, Bj{\"o}rn and Gamalielsson, Jonas},
  journal={Universal Access in the Information Society},
  volume={24},
  number={4},
  pages={3615--3637},
  year={2025},
  publisher={Springer}
}

@inproceedings{pool2023accessibility,
  title={Accessibility metatesting: comparing nine testing tools},
  author={Pool, Jonathan Robert},
  booktitle={Proceedings of the 20th International Web for All Conference},
  pages={1--4},
  year={2023}
}

@online{w3c_wcag22_2024,
  author = {{World Wide Web Consortium}},
  title = {Web Content Accessibility Guidelines (WCAG) 2.2},
  year = {2024},
  month = dec,
  day = {12},
  url = {https://www.w3.org/TR/2024/REC-WCAG22-20241212/}
}

@inproceedings{10.1145/355460.355549,
author = {Sullivan, Terry and Matson, Rebecca},
title = {Barriers to use: usability and content accessibility on the Web's most popular sites},
year = {2000},
isbn = {1581133146},
publisher = {Association for Computing Machinery},
address = {New York, NY, USA},
url = {https://doi.org/10.1145/355460.355549},
doi = {10.1145/355460.355549},
booktitle = {Proceedings on the 2000 Conference on Universal Usability},
pages = {139–144},
numpages = {6},
location = {Arlington, Virginia, USA},
series = {CUU '00}
}

@inproceedings{mack2021we,
  title={What do we mean by “accessibility research”? A literature survey of accessibility papers in CHI and ASSETS from 1994 to 2019},
  author={Mack, Kelly and McDonnell, Emma and Jain, Dhruv and Lu Wang, Lucy and E. Froehlich, Jon and Findlater, Leah},
  booktitle={Proceedings of the 2021 CHI Conference on Human Factors in Computing Systems},
  pages={1--18},
  year={2021}
}

@article{hong2007evaluating,
  title={Evaluating government website accessibility: Software tool vs human experts},
  author={Hong, Soongoo and Katerattanakul, Pairin and Lee, Dae-hyung},
  journal={Management Research News},
  volume={31},
  number={1},
  pages={27--40},
  year={2007},
  publisher={Emerald Group Publishing Limited}
}

@inproceedings{10.1145/3377811.3380392,
author = {Alshayban, Abdulaziz and Ahmed, Iftekhar and Malek, Sam},
title = {Accessibility issues in Android apps: state of affairs, sentiments, and ways forward},
year = {2020},
isbn = {9781450371216},
publisher = {Association for Computing Machinery},
address = {New York, NY, USA},
url = {https://doi.org/10.1145/3377811.3380392},
doi = {10.1145/3377811.3380392},
booktitle = {Proceedings of the ACM/IEEE 42nd International Conference on Software Engineering},
pages = {1323–1334},
numpages = {12},
location = {Seoul, South Korea},
series = {ICSE '20}
}

@article{ismail2022web,
  title={Web accessibility investigation and identification of major issues of higher education websites with statistical measures: A case study of college websites},
  author={Ismail, Abid and Kuppusamy, KS},
  journal={Journal of King Saud University-Computer and Information Sciences},
  volume={34},
  number={3},
  pages={901--911},
  year={2022},
  publisher={Elsevier}
}

@article{asok2025digital,
  title={Digital inclusion in higher education: A web content accessibility evaluation of best Asian university library websites},
  author={Asok, AR Arya and Rekha, RV},
  journal={The Journal of Academic Librarianship},
  volume={51},
  number={5},
  pages={103120},
  year={2025},
  publisher={Elsevier}
}

@misc{google2025lighthouse,
  author={{Google Chrome Developers}},
  title={Lighthouse Accessibility Score},
  year={2025},
  url={https://developer.chrome.com/docs/lighthouse/accessibility/scoring},
  note={Accessed July 27, 2026}
}

@article{singh2025openai,
  title={Openai gpt-5 system card},
  author={Singh, Aaditya and Fry, Adam and Perelman, Adam and Tart, Adam and Ganesh, Adi and El-Kishky, Ahmed and McLaughlin, Aidan and Low, Aiden and Ostrow, AJ and Ananthram, Akhila and others},
  journal={arXiv preprint arXiv:2601.03267},
  year={2025}
}
\bibliographystyle{colm2026_conference}

\clearpage

\appendix

\section{UIReq-6.8K: Application Domains}
\label{app:application_domain}

The training dataset covers 68 diverse application categories. Examples include:

\begin{tcolorbox}[title=Training Set Application Domains]
\begin{multicols}{2}
\begin{itemize}
    \item Business \& Enterprise
    \item Health \& Wellness
    \item Education \& E-Learning
    \item Data \& Analytics
    \item Communication \& Social
    \item E-Commerce \& Retail
    \item Finance \& FinTech
    \item Real Estate \& Property
    \item Media \& Entertainment
    \item Food \& Beverage
    \item Travel \& Hospitality
    \item Developer Tools \& Technology
    \item Science \& Research
    \item Legal \& Compliance
    \item Automotive \& Mobility
    \item Government \& Public Services
    \item Environment \& Sustainability
    \item Security \& Identity
    \item Non-Profit \& Social Impact
    \item AI \& Machine Learning
    \item Books \& Reference
    \item Comics
    \item Dating
    \item Entertainment
    \item Events
    \item Finance
    \item Food \& Drink
    \item Health \& Fitness
    \item House \& Home
    \item Libraries \& Demo
    \item Lifestyle
    \item Maps \& Navigation
    \item Medical
    \item Music \& Audio
    \item News \& Magazines
    \item Parenting
    \item Personalization
    \item Photography
    \item Productivity
    \item Shopping
    \item Social
    \item Sports
    \item Tools
    \item Travel \& Local
    \item Video Players \& Editors
    \item Weather
    \item Auto \& Vehicles
    \item Beauty
    \item Art \& Design
    \item Board
    \item Card
    \item Casino
    \item Casual
    \item Educational (Games)
    \item Music (Games)
    \item Puzzle
    \item Racing
    \item Role Playing
    \item Simulation
    \item Sports (Games)
    \item Strategy
    \item Trivia
    \item Word
    \item Augmented Reality
    \item Developer Tools
    \item Magazines \& Newspapers
    \item Utilities
    \item Graphics \& Design
\end{itemize}
\end{multicols}
\end{tcolorbox}

\vfill
\clearpage

\section{RealUIReq-300 Domain Distribution}

\begin{tcolorbox}[title=61 Domain Distribution of RealUIReq-300 Dataset]
\begin{multicols}{2}
\begin{description}
    \item[Education] 30
    \item[E-commerce] 19
    \item[Non-profit / Humanitarian] 13
    \item[Social media] 12
    \item[Web development] 10
    \item[Entertainment / Media] 9
    \item[Business consulting] 6
    \item[Academic publishing] 6
    \item[Personal development] 5
    \item[Finance / Investing] 4
    \item[Government services] 4
    \item[Environmental activism] 4
    \item[Community forum] 4
    \item[B2B marketing / Business services] 4
    \item[Business Collaboration] 3
    \item[Medical research] 3
    \item[Financial Services] 3
    \item[Crowdfunding] 2
    \item[Music streaming] 2
    \item[Wedding planning] 2
    \item[Creative networking] 2
    \item[Parenting / Community] 2
    \item[Advocacy / Activism] 2
    \item[Cloud computing / AI] 2
    \item[Search engine] 1
    \item[Culture / History] 1
    \item[Freelance marketplace] 1
    \item[Service industry] 1
    \item[Microfinance] 1
    \item[Space exploration] 1
    \item[Religious / Charitable Org] 1
    \item[Technology / Software] 23
    \item[News / Media] 16
    \item[Software development] 12
    \item[Tourism] 10
    \item[SaaS] 9
    \item[Entertainment ticketing] 8
    \item[Professional services] 6
    \item[Food / Cooking] 5
    \item[Travel / Lifestyle] 5
    \item[Tech discovery platforms] 4
    \item[Academic research] 4
    \item[Health and wellness] 4
    \item[Art and design] 4
    \item[Creative / Portfolio] 3
    \item[Content publishing] 3
    \item[Personal blog] 3
    \item[Consumer electronics] 2
    \item[Adventure / Outdoor sports] 2
    \item[Education / Interactive media] 2
    \item[Events / Entertainment] 2
    \item[Film / Photography] 2
    \item[Job board / Recruitment] 2
    \item[Healthcare tech] 2
    \item[Retail / Fashion] 1
    \item[Dev social] 1
    \item[Women empowerment] 1
    \item[Science museum] 1
    \item[Gaming / Media] 1
    \item[Cloud storage] 1
    \item[Humor / Lifestyle] 1
\end{description}
\end{multicols}
\end{tcolorbox}

\section{Training Configuration}
\label{app:train_config}
We trained \textit{Qwen/Qwen2.5-Coder-7B-Instruct} on 4 NVIDIA RTX PRO 6000 Blackwell GPUs (96GB VRAM each) using GRPO with vLLM-based sampling and reward modeling. Training was conducted in bfloat16 mixed-precision with gradient checkpointing enabled. Each prompt was expanded into $G=8$ sampled completions, with a per-device batch size of 2 and gradient accumulation set to 12. To stabilize optimization, KL divergence regularization was applied with $\beta = 0.001$. Below is a summary of the key configurations used.

\begin{table}[h]
\centering
\small
\setlength{\tabcolsep}{6pt}
\renewcommand{\arraystretch}{1.2}
\begin{tabular}{p{0.3\columnwidth}|p{0.5\columnwidth}}
\textbf{Component} & \textbf{Configuration} \\
\hline
Framework & HF Transformers; TRL (GRPOTrainer); Accelerate; DeepSpeed \\
\hline
Learning rate & $5 \times 10^{-5}$ \\
\hline
Optimizer & Adam ($\beta_1{=}0.9$, $\beta_2{=}0.99$, $\epsilon{=}10^{-8}$) \\
\hline
Weight decay & 0.1 \\
\hline
LR scheduler & Cosine (warmup ratio 0.01) \\
\hline
Gradient accumulation & 12 \\
\hline
Batch size & 96 \\
\hline
Epochs & 1 \\
\hline
Gradient checkpointing & Enabled \\
\hline
Precision & bfloat16 \\
\hline
Loss type & GRPO \\
\hline
KL $\beta$ & 0.001 \\
\hline
PEFT & Enabled \\
\hline
PEFT Defaults & LoRA r=16, alpha=32, dropout=0.05 \\
\end{tabular}
\caption{Optimization and training parameters.}
\label{tab:train_params}
\end{table}

\begin{table}[h]
\centering
\small
\setlength{\tabcolsep}{6pt}
\renewcommand{\arraystretch}{1.2}
\begin{tabular}{p{0.3\columnwidth}|p{0.5\columnwidth}}
\textbf{Parameter} & \textbf{Value} \\
\hline
Engine & vLLM (colocated mode) \\
\hline
GPU memory utilization & 60\% \\
\hline
Min-$p$ & 0.1 \\
\hline
Top-$p$ & 1.0 \\
\hline
Top-$k$ & $-1$ \\
\hline
Temperature & 0.7 \\
\hline
Repetition penalty & 1.1 \\
\hline
Stop sequence & \texttt{</answer>} \\
\hline
Max completion length & 3072 \\
\hline
Max prompt length & 1024 \\
\hline
Number of generations & 8 \\
\hline
Seed & 3407 \\
\end{tabular}
\caption{GRPO sampling parameters.}
\label{tab:sampling_params}
\end{table}

\begin{figure}[h]
  \centering
  \includegraphics[width=0.7\columnwidth]{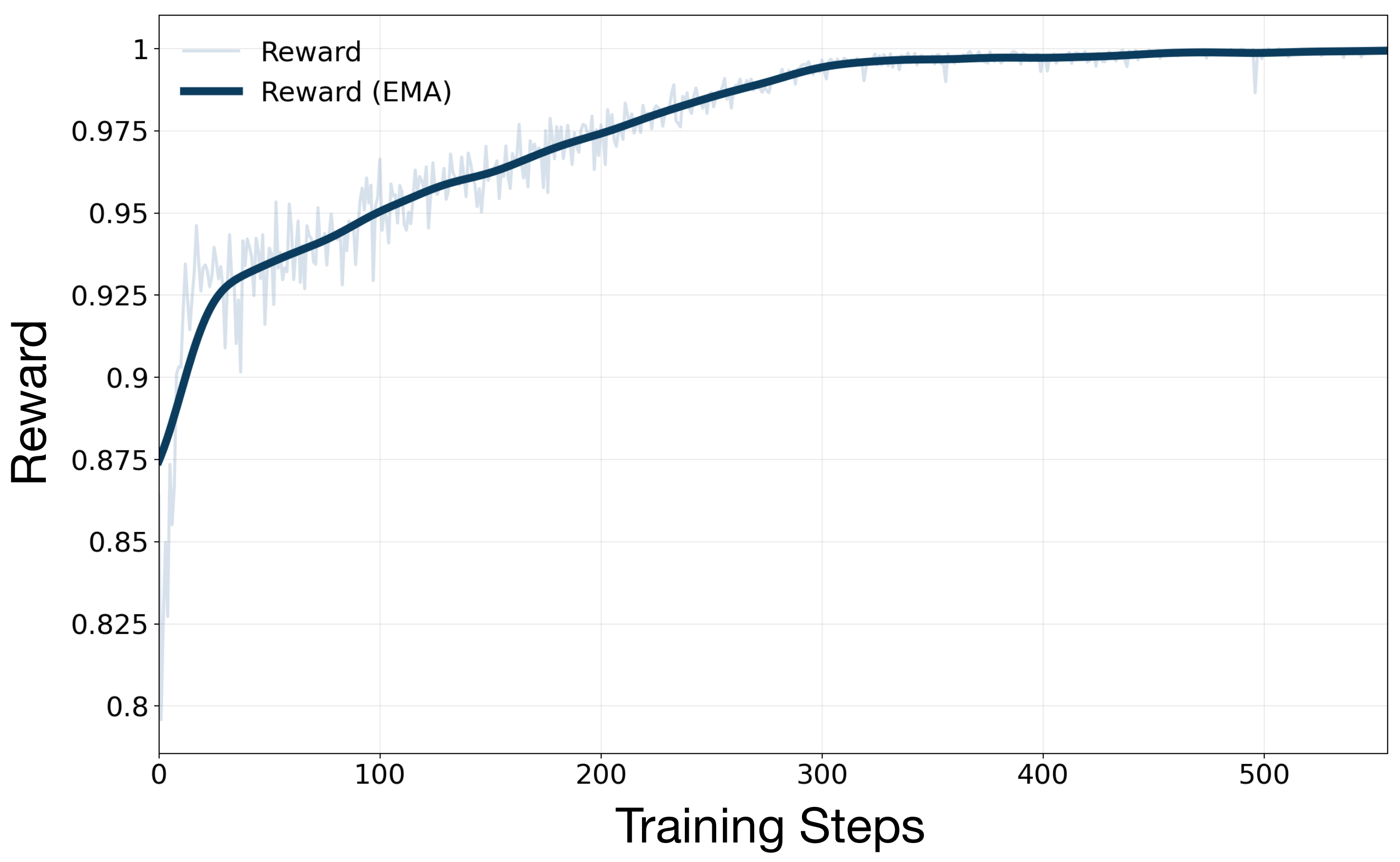}
  \caption{\textbf{Web Accessibility Reward curve throughout training.} The curve shows a steady increase indicating reduced WCAG violation occurrences over time.}
  \label{fig:accessibility_reward}
\end{figure}

\subsection{Training-Time Severity Distribution}
\label{app:training_severity}

We audit base-model generations on all UIReq-6.8K training prompts with Axe-core. 

\begin{table}[h]
\centering
\footnotesize
\setlength{\tabcolsep}{8pt}
\renewcommand{\arraystretch}{1.05}
\begin{tabular}{lr}
\toprule
\textbf{Severity} & \textbf{Affected Nodes}\\
\midrule
Minor & 121 \\
Moderate & 17,581 \\
Serious & 12,138\\
Critical & 672 \\
\bottomrule
\end{tabular}
\caption{\textbf{Initial training-time severity distribution.} Axe-core affected-node counts from base-model generations on UIReq-6.8K before A11yn optimization.}
\label{tab:training_severity}
\end{table}

Critical rules are sparse in the base model's training-time violation profile. This helps explain why A11yn reduces Critical affected nodes from 47 to 25 but trails Best-of-N at 12, especially for \texttt{button-name} and \texttt{select-name}, which appear only 135 and 155 times in the training audit. Thus, A11yn's strength is broad aggregate violation reduction, where targeted prompt augmentation on training set may further improve some underrepresented rules.

\section{Data Annotation of RealUIReq-300}
\label{data_annotation_detail}

The construction of RealUIReq-300 involved human-in-the-loop interventions to ensure semantic accuracy and realism of evaluation requests. Human contributors consisted of the paper authors. Human involvement occurred at three stages:

\begin{itemize}
    \item \textbf{UI Captioning Validation.}  
LLM-generated metadata extracted from webpage screenshots (e.g., UI purpose, page type, domain, and key components) was manually reviewed to verify that notable and functionally important UI elements were correctly captured. Among the set, 14\% (43/300) had low-resolution or unclear screenshots, for which the authors manually visited the corresponding webpages and rewrote the prospective UI request with explicit specifications (UI purpose, page type, domain, key components).

    \item \textbf{UI Request Draft Editing.}  
Initial request drafts were edited to improve clarity, specificity, and naturalness, resolving vague phrasing and ensuring alignment with the underlying UI content. For this process, tone biases (e.g., "Make a page ...") were spotted, and the annotators manually diversified the tone for all 300 requests while preserving the original specifications, so that the final benchmark reflects human-reviewed tone rather than a model's preferences.

    \item \textbf{Consistency and Coherence Checking.}  
Final requests were reviewed to ensure consistency and semantic coherence between the request texts and the corresponding real-world UIs, including alignment with the intended UI purpose, structure, and static web UI scope.

\end{itemize}

\section{Human Evaluation of Visual and Semantic Quality}
\label{app:human_evaluation}

We conduct a pairwise human evaluation to assess whether A11yn's accessibility improvements preserve the visual and semantic quality of generated UIs relative to the base model. Five recruited evaluators provide a total of 1,330 judgments. For each judgment, an evaluator is shown the user-request metadata and two rendered UIs, one generated by the base model and one generated by A11yn. The left--right position of the two UIs is randomized. Evaluators select one of five responses: strongly prefer the left UI, weakly prefer the left UI, similar, weakly prefer the right UI, or strongly prefer the right UI. We aggregate the responses after resolving the randomized positions to the corresponding model identities.

\begin{table}[h]
\centering
\small
\setlength{\tabcolsep}{6pt}
\renewcommand{\arraystretch}{1.1}
\begin{tabular}{lrr}
\toprule
\textbf{Human Judgment} & \textbf{Count} & \textbf{Rate} \\
\midrule
Strong preference for A11yn & 182 & 13.7\% \\
Weak preference for A11yn & 340 & 25.6\% \\
Similar & 523 & 39.3\% \\
Weak preference for Base & 196 & 14.7\% \\
Strong preference for Base & 89 & 6.7\% \\
\bottomrule
\end{tabular}
\caption{\textbf{Pairwise human evaluation of A11yn and the base model.} Counts and rates are aggregated over 1,330 judgments after resolving randomized left--right positions.}
\label{tab:human_evaluation}
\end{table}

As shown in Table~\ref{tab:human_evaluation}, the most common response is that the two outputs are similar (39.3\%). A11yn is preferred in another 39.3\% of all judgments, compared with 21.4\% for the base model. Among non-tie judgments, 64.7\% favor A11yn. Overall, the results indicate that accessibility alignment does not reduce perceived visual or semantic quality and may improve it for some requests.

\section{FeedA11y Baseline Implementation Details}
\label{app:feeda11y_impl}

We implement the FeedA11y baseline by following its three-stage loop: \textit{Generation}, \textit{Evaluation}, and \textit{ReAct Refinement}. Each stage is run by an independent agent, all backed by Qwen2.5-Coder-7B-Instruct to keep the backbone consistent with our primary controlled comparisons. During refinement, we retain the original FeedA11y editing constraints (e.g., ``fix only where necessary'' and ``preserve structure/functionality'') to avoid unnecessary rewrites.

\paragraph{Stage-wise procedure.}
Given a user request, the baseline first generates an initial HTML/CSS candidate (Generation). The candidate is then audited to produce accessibility findings (Evaluation). The ReAct agent consumes the code and the violation report, proposes targeted edits, and outputs a refined version (ReAct Refinement). This evaluation--refinement cycle is repeated with strict stopping criteria.

\paragraph{Stopping criteria.}
To prevent over-editing and infinite refinement loops, we use three deterministic stopping rules:
\begin{enumerate}
  \item terminate immediately when no accessibility violations are detected;
  \item stop early when refined code is identical to the previous iteration (code convergence);
  \item cap refinement at a maximum of three rounds.
\end{enumerate}

\paragraph{Differences from the original FeedA11y setup.}
Our implementation differs from the original FeedA11y in two aspects:
\begin{enumerate}
  \item \textbf{Prompt instantiation.} The original formulation is summary-to-code reconstruction, whereas our benchmark is request-to-code generation. We therefore adapt prompt wording to the request-style task while preserving the same three-stage control flow.
  \item \textbf{Auditing signal in Evaluation.} The original FeedA11y implementation generates accessibility reports via an LLM. In our pipeline, Evaluation uses auditor-grounded Axe-core output (e.g., rule IDs and descriptions) as the refinement signal for fairness.
\end{enumerate}

This design keeps the overall FeedA11y workflow unchanged while aligning the auditing interface with our auditor-based evaluation protocol. It reduces confounds from model-dependent violation reporting and supports fairer comparison under a shared WCAG auditing signal.

\section{Limitations and Future Work}
\paragraph{Reliance on Accessibility Audits.}
Following prior accessibility-technology studies, we operationalize web accessibility using automated auditors (Axe-core and AChecker) rather than direct human-centered assessment. Fully capturing the diverse, context-dependent accessibility needs of users through human evaluation remains an open challenge in HCI and is left for future work.

\paragraph{Scope of HTML-based Web UIs.} The current study focuses on HTML-based web UI generation, which covers a broad and practical class of real-world interfaces. Nevertheless, extending accessibility-aligned optimization to additional frontend frameworks and UI technologies remains an important direction as accessibility requirements may vary across frontend platforms.

\paragraph{Targeting of Static UI.} Our analysis considers static UI renderings derived from screenshots and does not explicitly model interactive behaviors. As some accessibility challenges arise from runtime interactions, future work may explore dynamic and interactive web interfaces, including accessibility considerations under UI state changes and user interactions.

\section{Prompt Details}
\label{app:prompt_detail}
We provide the details of the prompts used.

\subsection{Prompts for Dataset Synthesis}
\begin{tcolorbox}[title=Prompt for Generating Training Data,label={box:data},width=\textwidth]
\begin{verbatim}
Your task is to generate 100 different possible UI request 
queries for a certain application domain category.

The application category is:
{category}
Requirements for queries:
1. Each query should include different page types possible 
    within the app in the application domain
2. Each queries should be specific about widget requirements 
    and context rich (side navigation, collapsible menus, etc.)
3. Each query must be different from the others
4. Queries should be realistic and natural like real user 
    requests
5. Queries should be in some length and semantically rich, 
    not just a few words
6. Queries should have solid use case or purpose, not random 
    requests
7. Do not consider interactivity or animations or hovering 
    effects focus on static UI elements
8. Queries require some ambience or style requests like 
    color scheme, typography, etc.
\end{verbatim}
\end{tcolorbox}

\vfill
\clearpage

\begin{tcolorbox}[title=Prompt for Captioning metadata for evaluation set curation,label={box:captioning},width=\columnwidth]
\begin{verbatim}
Caption given UI screenshot with
1. Main purpose and intent of the UI, regarding the target audience and
actual use case.
2. Page type of the UI, such as a landing page, blog post, product page, etc. 
(in short-answer format)
3. Domain of the UI, such as e-commerce, social media, education, etc. 
(in short-answer format)
4. Top 5 important visual elements in the UI design, that are crucial for more 
user engagement and usability. But you should EXCLUDE any elements about the 
EXACT IMAGES in the UI. 
Answer in such format:
1. <Purpose/Intent>
2. <Page Type>
3. <Domain>
4. (a) <Element (a)>
(b) <Element (b)>
(c) <Element (c)>
...
\end{verbatim}
\end{tcolorbox}

\begin{tcolorbox}[title=Prompt for Generating RealUIReq-300 evaluation set queries, label={box:query_gen}, width=\columnwidth]
\begin{verbatim}
SYSTEM PROMPT:
You are a helpful assistant that generates realistic user 
requests for web UI development.

USER PROMPT:
Based on the following web page specifications, generate a
user request that mentions EVERY DETAIL provided, including 
the purpose, page type, domain, and all listed components.

Purpose: {purpose}
Page Type: {page_type}
Domain: {application_domain}
Required Components: {required_components}

The request should be 3-5 sentences long and sound realistic.
\end{verbatim}
\end{tcolorbox}

\vfill
\clearpage

\subsection{Prompts for Inference and Evaluation}
\begin{tcolorbox}[title=Prompt for Inference of models for web UI generation, label={box:inference}, width=\textwidth]
\begin{verbatim}
You are an expert UI designer assistant.
You should plan the design based on the user request. Show the plan in the 
`<think>` tag.
    - You must think about the html structure and widgets needed to fulfill 
      the user request.
    - You must think about the Tailwind CSS classes to use for styling.
Then, you should generate a complete HTML document that includes:
    - A `<head>` section with a `<meta charset="UTF-8">` tag
    - A `<meta name="viewport" content="width=device-width, initial-scale=
      1.0">` tag
    - A proper tailwind css link tag to load Tailwind CSS from CDN
    - A `<body>` section that contains the complete HTML structure and content
The HTML document should be visually appealing, well-structured, and content
/semantically-rich.
You must strictly follow the output format shown below:
<think>
...
</think>

<answer>
```html
<html>
<head>
    <meta charset="UTF-8" />
    <meta name="viewport" content="width=device-width, initial-scale=1.0" />
    ...
</head>
<body>
...
</body>
</html>
```
</answer>

User: {user_request}
Assistant:
\end{verbatim}
\end{tcolorbox}

\vfill
\clearpage

\begin{tcolorbox}[title=WebGen-Bench Appearance Score Evaluation
Prompt~\citep{lu2025webgenbenchevaluatingllmsgenerating},label={box:webgenbench},width=\textwidth]
\begin{verbatim}
Instruction: You are tasked with evaluating the functional design of a webpage 
that had been constructed based on the following instruction: {instruction}

Grade the webpage's appearance on a scale of 1 to 5 (5 being highest), 
considering the following criteria:
   - Successful Rendering: Does the webpage render correctly without visual
     errors? Are colors, fonts, and components displayed as specified?
   - Content Relevance: Does the design align with the webpage's purpose
     and user requirements? Are elements (e.g., search bars, report formats)
     logically placed and functional?
   - Layout Harmony: Is the arrangement of components (text, images, buttons) 
     balanced, intuitive, and clutter-free?
   - Modernness & Beauty: Does the design follow contemporary trends (e.g., 
     minimalism, responsive layouts)? Are colors, typography, and visual 
     hierarchy aesthetically pleasing?
Grading Scale:
   - 1 (Poor): Major rendering issues (e.g., broken layouts, incorrect colors). 
     Content is irrelevant or missing. Layout is chaotic. Design is outdated or 
    visually unappealing.
   - 2 (Below Average): Partial rendering with noticeable errors. Content is 
     partially relevant but poorly organized. Layout lacks consistency. Design 
    is basic or uninspired.
   - 3 (Average): Mostly rendered correctly with minor flaws. Content is 
     relevant but lacks polish. Layout is functional but unremarkable. Design is 
    clean but lacks modern flair.
   - 4 (Good): Rendered well with no major errors. Content is relevant and 
     logically organized. Layout is harmonious and user-friendly. Design is 
    modern and visually appealing.
   - 5 (Excellent): Flawless rendering. Content is highly relevant, intuitive, 
     and tailored to user needs. Layout is polished, responsive, and innovative. 
    Design is cutting-edge, beautiful, and memorable.
     
Task: Review the provided screenshot(s) of the webpage. Provide a detailed 
analysis and then assign a grade (from 1 to 5) based on your analysis. Highlight
strengths, weaknesses, and how well the design adheres to the specifications. 

IMPORTANT: Please end your response with a clear grade in the format "Grade: X" 
where X is a number from 1 to 5. 

Your Response Format: 
  Analysis: [from 2 to 4 paragraphs addressing all criteria, referencing the 
  instruction]
  Grade: [from 1 to 5]
  Your Response:
\end{verbatim}
\end{tcolorbox}

\section{Additional Evaluation Statistics}
\label{app:additional_stats}

We report complementary diagnostics beyond aggregate metrics in Table~\ref{tab:quant_results}. Specifically, we provide UI domain-level accessibility outcomes (Table~\ref{tab:normalized_domain_coverage}), page-level pass counts per auditor (Table~\ref{tab:core_additional_stats}), and per-request pairwise outcomes (Table~\ref{tab:per_request_wins}).

\subsection{Normalized Domain Coverage}
\label{app:normalized_domain_coverage}

Table~\ref{tab:normalized_domain_coverage} follows the exact 61 normalized domain bins and counts from the RealUIReq-300 domain-distribution chart. For each method and domain, we report: (1) \textbf{Count}, the number of pages in that domain that fail at least one auditor; (2) \textbf{$IR_{\text{AChecker}}$}; and (3) \textbf{$IR_{\text{Axe}}$}. Lower is better for all metrics. Boldface marks the lowest value within each row (Count, $IR_{\text{AChecker}}$, $IR_{\text{Axe}}$); ties are all bolded.

\begin{table*}[h]
\centering
\scriptsize
\setlength{\tabcolsep}{1pt}
\renewcommand{\arraystretch}{1.03}
\resizebox{\textwidth}{!}{
\begin{tabular}{lccccccccccccc}
\toprule
\multirow{2}{*}{\textbf{Normalized Domain Family}} & \multirow{2}{*}{\textbf{N}} & \multicolumn{3}{c}{\textbf{A11yn}} & \multicolumn{3}{c}{\textbf{Base}} & \multicolumn{3}{c}{\textbf{FeedA11y}} & \multicolumn{3}{c}{\textbf{Best-of-8}} \\
\cmidrule(lr){3-5}\cmidrule(lr){6-8}\cmidrule(lr){9-11}\cmidrule(lr){12-14}
& & \textbf{Count} & \textbf{$IR_{\text{AChecker}}$} & \textbf{$IR_{\text{Axe}}$} & \textbf{Count} & \textbf{$IR_{\text{AChecker}}$} & \textbf{$IR_{\text{Axe}}$} & \textbf{Count} & \textbf{$IR_{\text{AChecker}}$} & \textbf{$IR_{\text{Axe}}$} & \textbf{Count} & \textbf{$IR_{\text{AChecker}}$} & \textbf{$IR_{\text{Axe}}$} \\
\midrule
Education & 30 & \textbf{17} & \textbf{0.27} & \textbf{0.06} & 30 & 0.64 & 0.39 & 29 & 0.55 & 0.38 & 21 & 0.41 & 0.15 \\
E-commerce & 19 & \textbf{10} & 0.31 & \textbf{0.08} & 18 & 0.56 & 0.39 & 19 & 0.68 & 0.45 & 16 & \textbf{0.21} & 0.11 \\
Non-profit / Humanitarian & 13 & \textbf{5} & \textbf{0.21} & \textbf{0.07} & 13 & 0.70 & 0.65 & 13 & 0.45 & 0.28 & 10 & 0.45 & 0.16 \\
Social media & 12 & \textbf{7} & \textbf{0.24} & \textbf{0.04} & 12 & 0.67 & 0.43 & 12 & 0.60 & 0.37 & 12 & 0.41 & 0.16 \\
Web development & 10 & \textbf{3} & 0.46 & \textbf{0.05} & 10 & 0.63 & 0.31 & 9 & 0.56 & 0.22 & 9 & \textbf{0.45} & 0.13 \\
Entertainment / Media & 9 & \textbf{3} & \textbf{0.14} & \textbf{0.02} & 9 & 0.58 & 0.26 & 9 & 0.47 & 0.31 & 5 & 0.36 & 0.09 \\
Business consulting & 6 & \textbf{2} & \textbf{0.08} & \textbf{0.03} & 6 & 0.56 & 0.34 & 6 & 0.41 & 0.18 & 4 & 0.41 & 0.13 \\
Academic publishing & 6 & \textbf{4} & \textbf{0.30} & \textbf{0.06} & 5 & 0.59 & 0.37 & 5 & 0.49 & 0.15 & 5 & 0.41 & 0.13 \\
Personal development & 5 & \textbf{1} & 0.23 & \textbf{0.00} & 4 & 0.45 & 0.59 & 4 & 0.32 & 0.19 & 2 & \textbf{0.20} & 0.04 \\
Finance / Investing & 4 & 4 & 0.62 & 0.12 & 4 & 0.52 & 0.25 & 4 & 0.58 & 0.29 & \textbf{3} & \textbf{0.15} & \textbf{0.07} \\
Government services & 4 & \textbf{2} & 0.33 & 0.08 & 4 & 0.79 & 0.65 & 3 & 0.53 & 0.18 & \textbf{2} & \textbf{0.13} & \textbf{0.04} \\
Environmental activism & 4 & \textbf{3} & \textbf{0.40} & \textbf{0.11} & \textbf{3} & 0.67 & 0.34 & \textbf{3} & 0.53 & 0.68 & \textbf{3} & 0.48 & 0.22 \\
Community forum & 4 & 3 & 0.33 & 0.05 & 4 & 0.48 & 0.28 & 4 & 0.65 & 0.38 & \textbf{1} & \textbf{0.10} & \textbf{0.03} \\
B2B marketing / Business services & 4 & \textbf{1} & \textbf{0.21} & \textbf{0.00} & 4 & 0.68 & 0.34 & 3 & 0.50 & 0.21 & 3 & 0.48 & 0.14 \\
Business Collaboration & 3 & \textbf{1} & \textbf{0.43} & \textbf{0.00} & 3 & 0.68 & 0.87 & 3 & 0.47 & 0.32 & 2 & 0.50 & 0.24 \\
Medical research & 3 & \textbf{0} & \textbf{0.00} & \textbf{0.00} & 3 & 0.77 & 0.36 & 3 & 0.45 & 0.36 & 3 & 0.20 & 0.10 \\
Financial Services & 3 & \textbf{1} & \textbf{0.06} & \textbf{0.03} & 3 & 0.65 & 0.35 & 3 & 0.60 & 0.28 & \textbf{1} & 0.36 & 0.07 \\
Crowdfunding & 2 & \textbf{1} & \textbf{0.06} & \textbf{0.03} & 2 & 0.63 & 0.34 & 2 & 0.54 & 0.31 & 2 & 0.72 & 0.23 \\
Music streaming & 2 & \textbf{2} & 0.22 & \textbf{0.12} & \textbf{2} & 0.35 & 0.20 & \textbf{2} & 0.61 & 0.29 & \textbf{2} & \textbf{0.21} & \textbf{0.12} \\
Wedding planning & 2 & \textbf{2} & \textbf{0.46} & \textbf{0.10} & \textbf{2} & 0.57 & 0.43 & \textbf{2} & 0.55 & 0.32 & \textbf{2} & 0.59 & 0.21 \\
Creative networking & 2 & \textbf{1} & 0.25 & 0.03 & 2 & 0.44 & 0.14 & 2 & 0.67 & 0.44 & \textbf{1} & \textbf{0.17} & \textbf{0.02} \\
Parenting / Community & 2 & \textbf{1} & 0.40 & 0.07 & 2 & 0.22 & 0.17 & 2 & 0.33 & 0.10 & \textbf{1} & \textbf{0.20} & \textbf{0.06} \\
Advocacy / Activism & 2 & \textbf{2} & 0.50 & \textbf{0.32} & \textbf{2} & 0.82 & 0.50 & \textbf{2} & 0.77 & 0.55 & \textbf{2} & \textbf{0.40} & 0.33 \\
Cloud computing / AI & 2 & \textbf{1} & 0.33 & \textbf{0.00} & 2 & 0.42 & 0.39 & 2 & \textbf{0.18} & 0.10 & 2 & 0.50 & 0.20 \\
Search engine & 1 & \textbf{0} & \textbf{0.00} & \textbf{0.00} & 1 & 0.40 & 0.15 & 1 & 0.50 & 0.25 & 1 & 0.33 & 0.11 \\
Culture / History & 1 & \textbf{1} & 0.75 & \textbf{0.20} & \textbf{1} & 0.88 & 0.24 & \textbf{1} & \textbf{0.50} & 0.27 & \textbf{1} & \textbf{0.50} & 0.24 \\
Freelance marketplace & 1 & \textbf{0} & \textbf{0.00} & \textbf{0.00} & 1 & 0.73 & 0.26 & 1 & 0.43 & 0.33 & 1 & 0.25 & 0.07 \\
Service industry & 1 & \textbf{0} & \textbf{0.00} & \textbf{0.00} & 1 & 0.64 & 0.60 & 1 & 0.61 & 0.77 & \textbf{0} & \textbf{0.00} & \textbf{0.00} \\
Microfinance & 1 & \textbf{0} & \textbf{0.00} & \textbf{0.00} & 1 & 0.50 & 0.75 & 1 & 0.64 & 0.68 & \textbf{0} & \textbf{0.00} & \textbf{0.00} \\
Space exploration & 1 & \textbf{0} & \textbf{0.00} & \textbf{0.00} & 1 & 0.33 & 0.17 & 1 & 0.88 & 1.33 & \textbf{0} & \textbf{0.00} & \textbf{0.00} \\
Religious / Charitable Org & 1 & \textbf{0} & \textbf{0.00} & \textbf{0.00} & 1 & 0.46 & 0.71 & 1 & 0.33 & 0.12 & 1 & 0.50 & 0.33 \\
Technology / Software & 23 & \textbf{11} & \textbf{0.27} & \textbf{0.03} & 22 & 0.62 & 0.31 & 23 & 0.52 & 0.31 & 20 & 0.49 & 0.10 \\
News / Media & 16 & \textbf{13} & 0.50 & \textbf{0.10} & 16 & 0.66 & 0.41 & 16 & 0.70 & 0.42 & 15 & \textbf{0.41} & 0.17 \\
Software development & 12 & \textbf{8} & \textbf{0.36} & \textbf{0.05} & 12 & 0.66 & 0.37 & 12 & 0.61 & 0.33 & 10 & 0.47 & 0.13 \\
Tourism & 10 & \textbf{6} & \textbf{0.23} & \textbf{0.05} & 10 & 0.68 & 0.58 & 10 & 0.52 & 0.37 & 10 & 0.56 & 0.23 \\
SaaS & 9 & \textbf{3} & \textbf{0.11} & \textbf{0.02} & 9 & 0.61 & 0.49 & 8 & 0.33 & 0.21 & 7 & 0.43 & 0.15 \\
Entertainment ticketing & 8 & \textbf{5} & \textbf{0.29} & \textbf{0.07} & 8 & 0.61 & 0.40 & 8 & 0.56 & 0.44 & 6 & 0.33 & 0.12 \\
Professional services & 6 & \textbf{2} & \textbf{0.24} & \textbf{0.09} & 6 & 0.56 & 0.34 & 5 & 0.35 & 0.15 & 5 & 0.54 & 0.36 \\
Food / Cooking & 5 & \textbf{1} & \textbf{0.10} & \textbf{0.01} & 4 & 0.65 & 0.43 & 4 & 0.23 & 0.09 & 4 & 0.14 & 0.02 \\
Travel / Lifestyle & 5 & \textbf{2} & \textbf{0.26} & \textbf{0.04} & 4 & 0.55 & 0.42 & 5 & 0.48 & 0.35 & 5 & 0.47 & 0.12 \\
Tech discovery platforms & 4 & \textbf{2} & \textbf{0.27} & \textbf{0.13} & 4 & 0.72 & 0.50 & 4 & 0.62 & 0.30 & 3 & 0.53 & 0.15 \\
Academic research & 4 & \textbf{3} & \textbf{0.41} & \textbf{0.06} & 4 & 0.68 & 0.38 & 4 & 0.61 & 0.30 & 4 & 0.50 & 0.16 \\
Health and wellness & 4 & \textbf{2} & \textbf{0.31} & \textbf{0.02} & 4 & 0.71 & 0.48 & 4 & 0.46 & 0.36 & 3 & 0.50 & 0.26 \\
Art and design & 4 & \textbf{1} & \textbf{0.03} & \textbf{0.00} & 3 & 0.50 & 0.25 & 4 & 0.59 & 0.37 & 3 & 0.22 & 0.05 \\
Creative / Portfolio & 3 & \textbf{1} & 0.25 & \textbf{0.02} & 3 & 0.64 & 0.66 & 3 & 0.60 & 0.49 & 2 & \textbf{0.08} & 0.05 \\
Content publishing & 3 & \textbf{1} & \textbf{0.00} & \textbf{0.02} & 3 & 0.63 & 0.36 & 3 & 0.51 & 0.30 & 3 & 0.27 & 0.13 \\
Personal blog & 3 & 2 & 0.44 & \textbf{0.02} & \textbf{1} & 0.43 & 0.08 & 3 & 0.62 & 0.29 & \textbf{1} & \textbf{0.33} & 0.04 \\
Consumer electronics & 2 & \textbf{1} & 0.33 & \textbf{0.06} & 2 & 0.61 & 0.31 & \textbf{1} & 0.60 & 0.18 & 2 & \textbf{0.11} & 0.14 \\
Adventure / Outdoor sports & 2 & \textbf{0} & \textbf{0.00} & \textbf{0.00} & 2 & 0.53 & 0.80 & 2 & 0.43 & 0.30 & 2 & 0.48 & 0.29 \\
Education / Interactive media & 2 & \textbf{1} & \textbf{0.14} & \textbf{0.02} & 2 & 0.43 & 0.33 & 2 & 0.50 & 0.18 & \textbf{1} & 0.25 & 0.04 \\
Film / Photography & 2 & 2 & 0.63 & 0.11 & 2 & 0.79 & 0.42 & 2 & 0.60 & 0.33 & \textbf{1} & \textbf{0.17} & \textbf{0.04} \\
Job board / Recruitment & 2 & \textbf{0} & \textbf{0.00} & \textbf{0.00} & 1 & 0.30 & 0.18 & 2 & 0.42 & 0.43 & 2 & 0.43 & 0.20 \\
Healthcare tech & 2 & \textbf{0} & \textbf{0.00} & \textbf{0.00} & 2 & 0.72 & 0.38 & 2 & 0.50 & 0.45 & 2 & 0.53 & 0.21 \\
Retail / Fashion & 1 & \textbf{0} & \textbf{0.00} & \textbf{0.00} & 1 & 0.82 & 1.08 & 1 & 0.72 & 0.92 & 1 & 0.53 & 0.36 \\
Dev social & 1 & \textbf{1} & 0.50 & 0.05 & \textbf{1} & \textbf{0.11} & 0.27 & \textbf{1} & 0.58 & 0.36 & \textbf{1} & 0.40 & \textbf{0.04} \\
Women empowerment & 1 & \textbf{0} & \textbf{0.00} & \textbf{0.00} & \textbf{0} & \textbf{0.00} & \textbf{0.00} & 1 & 0.13 & 0.06 & 1 & 0.33 & 0.19 \\
Science museum & 1 & \textbf{0} & \textbf{0.00} & \textbf{0.00} & 1 & 0.33 & 0.47 & 1 & 0.18 & 0.04 & 1 & 0.43 & 0.15 \\
Gaming / Media & 1 & \textbf{1} & 0.50 & 0.09 & \textbf{1} & \textbf{0.10} & 0.12 & \textbf{1} & 0.29 & 0.14 & \textbf{1} & 0.20 & \textbf{0.04} \\
Cloud storage & 1 & \textbf{0} & \textbf{0.00} & \textbf{0.00} & 1 & 0.83 & 1.84 & 1 & 0.29 & 0.21 & 1 & 0.62 & 0.38 \\
Humor / Lifestyle & 1 & \textbf{1} & 0.57 & 0.33 & \textbf{1} & 0.71 & \textbf{0.32} & \textbf{1} & \textbf{0.50} & \textbf{0.32} & \textbf{1} & 0.63 & 0.33 \\
Events / Entertainment & 2 & \textbf{1} & 0.60 & \textbf{0.10} & 2 & 0.69 & 0.39 & 2 & 0.49 & 0.21 & 2 & \textbf{0.45} & 0.25 \\
\bottomrule
\end{tabular}
}
\caption{\textbf{Domain-level accessibility performance on RealUIReq-300 (61 normalized domains).} For each method, columns report (i) Count: pages in that domain that fail at least one auditor, (ii) $IR_{\text{AChecker}}$, and (iii) $IR_{\text{Axe}}$. Bold marks the lowest value within each row for each metric block; ties are bolded. Lower is better.}
\label{tab:normalized_domain_coverage}
\end{table*}

\clearpage

\subsection{Page-level Pass Rates and Pairwise Win Rates}
To complement the domain-level results in Table~\ref{tab:normalized_domain_coverage}, we provide two prompt-level summaries on the same RealUIReq-300 benchmark. Table~\ref{tab:core_additional_stats} reports how many of the 300 generated pages have zero violations under Axe, AChecker, and both auditors jointly. Table~\ref{tab:per_request_wins} reports per-request head-to-head outcomes (W-L-T) between A11yn and each baseline, where a win indicates a lower page-level inaccessibility rate and ties are resolved by the weighted violation score.

\begin{table*}[h]
\centering
\small
\setlength{\tabcolsep}{8pt}
\renewcommand{\arraystretch}{1.1}
\begin{tabular}{lccc}
\toprule
\textbf{Method} & \textbf{\makecell[c]{Pages w/o\\Axe Violations}} & \textbf{\makecell[c]{Pages w/o\\AChecker Violations}} & \textbf{\makecell[c]{Pages w/o Both\\Violations}} \\
\midrule
Base & 18/300 & 19/300 & 12/300 \\
Best-of-8 & \underline{79/300} & \underline{86/300} & \underline{62/300} \\
FeedA11y & 16/300 & 17/300 & 11/300 \\
A11yn & \textbf{177/300} & \textbf{166/300} & \textbf{151/300} \\
\bottomrule
\end{tabular}
\caption{\textbf{Page-level pass counts on RealUIReq-300.} Each value is the number of pages out of 300 with zero violations under the specified auditor; \textit{Pages w/o Both Violations} requires passing both Axe and AChecker simultaneously.}
\label{tab:core_additional_stats}
\end{table*}

\begin{table}[h]
\centering
\small
\setlength{\tabcolsep}{8pt}
\renewcommand{\arraystretch}{1.1}
\begin{tabular}{lcc}
\toprule
\textbf{Comparison} & \textbf{W-L-T} & \textbf{Win Rate Excluding Ties} \\
\midrule
A11yn vs Base & 270-14-16 & 95.1\% \\
A11yn vs FeedA11y & 273-18-9 & 93.8\% \\
A11yn vs Best-of-8 & 190-52-58 & 78.5\% \\
\bottomrule
\end{tabular}
\caption{\textbf{Per-request pairwise outcomes (A11yn vs. each baseline) on RealUIReq-300.} W-L-T denotes wins/losses/ties across the same 300 prompts. A win means a lower page-level inaccessibility rate for that prompt.}
\label{tab:per_request_wins}
\end{table}

\section{Model Scaling with and without Accessibility-Aware Prompting}
\label{app:prompt_scaling}

To distinguish model capacity from the elicitation of accessibility objectives, we evaluate the Qwen2.5-Coder family under both the default UI-generation prompt and an accessibility-aware system prompt. Under default prompting, accessibility performance is non-monotonic. With accessibility-aware prompting, larger models generally obtain lower IR$_{\text{Axe}}$ and higher Lighthouse scores. These findings suggest that scale provides greater latent capacity to follow accessibility constraints, but such capacity is not reliably activated by default UI-generation instructions. Nevertheless, the prompted 32B model remains below A11yn (IR$_{\text{Axe}}$: 0.05, IR$_{\text{AChecker}}$: 0.26, Lighthouse: 97.68), supporting the benefit of our training-time alignment.

\begin{table*}[h]
\centering
\small
\setlength{\tabcolsep}{4pt}
\renewcommand{\arraystretch}{1.1}
\resizebox{\textwidth}{!}{%
\begin{tabular}{lcccccc}
\toprule
\multirow{2}{*}{\textbf{Model}} &
\multicolumn{3}{c}{\textbf{Default Prompt}} &
\multicolumn{3}{c}{\textbf{Accessibility-Aware Prompt}} \\
\cmidrule(lr){2-4}\cmidrule(lr){5-7}
& \textbf{IR$_{\text{Axe}}$ $\downarrow$} & \textbf{IR$_{\text{AChecker}}$ $\downarrow$} & \textbf{Lighthouse $\uparrow$}
& \textbf{IR$_{\text{Axe}}$ $\downarrow$} & \textbf{IR$_{\text{AChecker}}$ $\downarrow$} & \textbf{Lighthouse $\uparrow$} \\
\midrule
Qwen2.5-Coder-3B  & 0.56 & 0.67 & 71.91 & 0.43 & 0.62 & 80.35 \\
Qwen2.5-Coder-7B  & \textbf{0.40} & \textbf{0.62} & \textbf{90.88} & 0.29 & 0.57 & 90.11 \\
Qwen2.5-Coder-14B & 0.48 & 0.68 & 73.02 & 0.25 & 0.60 & 92.65 \\
Qwen2.5-Coder-32B & 0.42 & 0.68 & 81.07 & \textbf{0.18} & \textbf{0.49} & \textbf{93.65} \\
\bottomrule
\end{tabular}
}
\caption{\textbf{Model scaling under default and accessibility-aware prompting.} Explicitly eliciting accessibility requirements in the system prompt makes the scaling trend more consistent. Bold values mark the best result within each prompt setting.}
\label{tab:prompt_scaling}
\end{table*}
\clearpage

\section{Extended Best-of-N Scaling up to N=20}
\label{app:bon_extension}

\begin{figure}[h]
  \centering
  \includegraphics[width=\columnwidth]{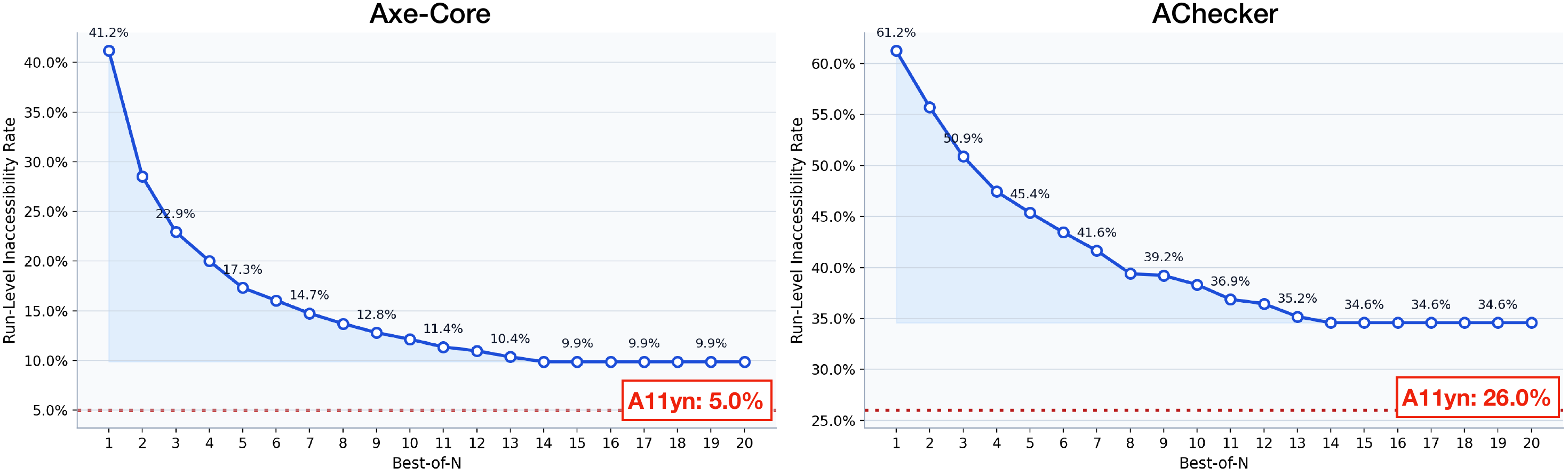}
\caption{\textbf{Extended Best-of-N Scaling under Axe-core and AChecker Evaluation.} As N increases, Best-of-N initially improves accessibility under both auditors, but the gains diminish and plateau within the tested range. Under Axe-core, the run-level inaccessibility rate decreases from 41.2\% at N=1 to 9.9\% at N=14 and remains at 9.9\% through N=20; under AChecker, it decreases from 61.2\% to 34.6\% at N=14 and likewise remains at 34.6\% through N=20. In both cases, the Inaccessibility Rate remains above A11yn at 5.0\% and 26.0\%, respectively, suggesting that A11yn's improvement is not reducible to test-time sampling and selection alone, and is instead consistent with RL-induced policy improvement.}
  \label{fig:RL_works}
\end{figure}

\clearpage

\section{WCAG Rule ID Mapping}
\label{app:wcag_rule_mapping}

\begin{table*}[h]
\centering
\small
\setlength{\tabcolsep}{4pt}
\renewcommand{\arraystretch}{1.12}
\begin{tabularx}{\textwidth}{p{0.24\textwidth} p{0.24\textwidth} X}
\toprule
\textbf{Plot Label} & \textbf{Rule ID} & \textbf{Appendix Description} \\
\midrule
Low text contrast & \texttt{color-contrast} & Elements do not meet minimum color contrast. \\
Unlandmarked content & \texttt{region} & Page content is not fully contained within landmark regions. \\
Nested main landmark & \makecell[l]{\texttt{landmark-main-is-top}\\\texttt{-level}} & The main landmark is nested inside another landmark. \\
Links without names & \texttt{link-name} & Links do not have discernible accessible text. \\
Ambiguous landmarks & \texttt{landmark-unique} & Landmarks do not have unique role or label combinations. \\
Missing or multiple main & \texttt{landmark-one-main} & The page does not have exactly one main landmark. \\
Multiple main landmarks & \makecell[l]{\texttt{landmark-no-duplicate-}\\\texttt{main}} & The page has more than one main landmark. \\
Missing main heading & \texttt{page-has-heading-one} & The page does not contain a level-one heading (H1). \\
Buttons without labels & \texttt{button-name} & Buttons do not have discernible accessible text. \\
Frames without titles & \texttt{frame-title} & Frames or iframes do not have an accessible name. \\
Selects without labels & \texttt{select-name} & Select elements do not have an accessible name. \\
Nested footer landmark & \makecell[l]{\texttt{landmark-}\\\texttt{contentinfo-}\\\texttt{is-top-level}} & The contentinfo landmark is nested inside another landmark. \\
Nested header landmark & \makecell[l]{\texttt{landmark-banner-}\\\texttt{is-top-level}} & The banner landmark is nested inside another landmark. \\
Broken heading order & \texttt{heading-order} & Heading levels increase by more than one level at a time. \\
Multiple header landmarks & \makecell[l]{\texttt{landmark-no-duplicate-}\\\texttt{banner}} & The page has more than one banner landmark. \\
Missing page title & \texttt{document-title} & The document is missing a meaningful title element. \\
Missing page language & \texttt{html-has-lang} & The page is missing a valid language attribute on the html element. \\
Repeated alt text & \texttt{image-redundant-alt} & Image alt text repeats adjacent visible text. \\
Nested sidebar landmark & \makecell[l]{\texttt{landmark-}\\\texttt{complementary-}\\\texttt{is-top-level}} & The complementary landmark is nested inside another landmark. \\
Missing input labels & \texttt{label} & Form controls are missing programmatically associated labels. \\
\bottomrule
\end{tabularx}
\caption{\textbf{WCAG rule mapping used in rule-level diagnostics.}}
\label{tab:wcag_rule_mapping}
\end{table*}

\end{document}